\documentclass[manuscript]{acmart}

\AtBeginDocument{%
  \providecommand\BibTeX{{%
    \normalfont B\kern-0.5em{\scshape i\kern-0.25em b}\kern-0.8em\TeX}}}

\setcopyright{acmcopyright}
\copyrightyear{2018}
\acmYear{2018}
\acmDOI{XXXXXXX.XXXXXXX}

\acmConference[Conference acronym 'XX]{Make sure to enter the correct
  conference title from your rights confirmation emai}{June 03--05,
  2018}{Woodstock, NY}
\acmPrice{15.00}
\acmISBN{978-1-4503-XXXX-X/18/06}

\usepackage{booktabs}
\usepackage{amsmath}
\usepackage{amsfonts}
\usepackage{multirow}
\usepackage{xspace}
\usepackage{subcaption}
\usepackage{graphicx}
\usepackage{wrapfig}
\usepackage{nicefrac}
\usepackage{soul}

\usepackage{afterpage}

\newcommand{\ie}{\emph{i.e.},\xspace}
\newcommand{\eg}{\emph{e.g.},\xspace}

\def\B#1{\mathbf #1}
\def\C#1{\mathcal #1}

\begin{document}

%%
%% The "title" command has an optional parameter,
%% allowing the author to define a "short title" to be used in page headers.
\title{Multimodal Pre-training for Sequential Recommendation via Contrastive Learning}

%%
%% The "author" command and its associated commands are used to define
%% the authors and their affiliations.
%% Of note is the shared affiliation of the first two authors, and the
%% "authornote" and "authornotemark" commands
%% used to denote shared contribution to the research.
% \author{Ben Trovato}
% \authornote{Both authors contributed equally to this research.}
% \email{trovato@corporation.com}
% \orcid{1234-5678-9012}
% \author{G.K.M. Tobin}
% \authornotemark[1]
% \email{webmaster@marysville-ohio.com}
% \affiliation{%
%   \institution{Institute for Clarity in Documentation}
%   \streetaddress{P.O. Box 1212}
%   \city{Dublin}
%   \state{Ohio}
%   \country{USA}
%   \postcode{43017-6221}
% }

% \author{Lars Th{\o}rv{\"a}ld}
% \affiliation{%
%   \institution{The Th{\o}rv{\"a}ld Group}
%   \streetaddress{1 Th{\o}rv{\"a}ld Circle}
%   \city{Hekla}
%   \country{Iceland}}
% \email{larst@affiliation.org}

% \author{Valerie B\'eranger}
% \affiliation{%
%   \institution{Inria Paris-Rocquencourt}
%   \city{Rocquencourt}
%   \country{France}
% }

% \author{Aparna Patel}
% \affiliation{%
%  \institution{Rajiv Gandhi University}
%  \streetaddress{Rono-Hills}
%  \city{Doimukh}
%  \state{Arunachal Pradesh}
%  \country{India}}

\author{Lingzi Zhang}
\affiliation{%
  \institution{Nanyang Technological University}
  \streetaddress{30 Shuangqing Rd}
  \city{Haidian Qu}
  \state{Beijing Shi}
  \country{Singapore}}

\author{Xin Zhou}
\affiliation{%
  \institution{Nanyang Technological University}
  \streetaddress{8600 Datapoint Drive}
  \city{San Antonio}
  \state{Texas}
  \country{Singapore}
  \postcode{78229}}
\email{cpalmer@prl.com}

\author{Zhiwei Zeng}
\affiliation{%
  \institution{Nanyang Technological University}
  \streetaddress{1 Th{\o}rv{\"a}ld Circle}
  \city{Hekla}
  \country{Singapore}}
\email{jsmith@affiliation.org}

\author{Zhiqi Shen}
\affiliation{%
  \institution{Nanyang Technological University}
  \city{New York}
  \country{Singapore}}
\email{jpkumquat@consortium.net}

%%
%% By default, the full list of authors will be used in the page
%% headers. Often, this list is too long, and will overlap
%% other information printed in the page headers. This command allows
%% the author to define a more concise list
%% of authors' names for this purpose.
% \renewcommand{\shortauthors}{Trovato and Tobin, et al.}

%%
%% The abstract is a short summary of the work to be presented in the
%% article.
\begin{abstract}
Sequential recommendation systems often suffer from data sparsity, leading to suboptimal performance. 
While multimodal content, such as images and text, has been utilized to mitigate this issue, its integration within sequential recommendation frameworks remains challenging. 
Current multimodal sequential recommendation models are often unable to effectively explore and capture correlations among behavior sequences of users and items across different modalities, either neglecting correlations among sequence representations or inadequately capturing associations between multimodal data and sequence data in their representations. 
To address this problem, we explore multimodal pre-training in the context of sequential recommendation, with the aim of enhancing fusion and utilization of multimodal
information.

We propose a novel \underline{M}ultimodal \underline{P}re-training for \underline{S}equential \underline{R}ecommendation (MP4SR) framework, which utilizes contrastive losses to capture the correlation among different modality sequences of users, as well as the correlation among different modality sequences of users and items.
MP4SR consists of three key components: 1) multimodal feature extraction, 2) a backbone network, Multimodal Mixup Sequence Encoder (M$^2$SE), and 3) pre-training tasks. After utilizing pre-trained encoders to generate initial multimodal features of items, M$^2$SE adopts a complementary sequence mixup strategy to fuse different modality sequences, and leverages contrastive learning to capture modality interactions at the sequence-to-sequence and sequence-to-item levels. Extensive experiments on four real-world datasets demonstrate that MP4SR outperforms state-of-the-art approaches in both normal and cold-start settings. 
We further highlight the efficacy of incorporating multimodal pre-training in sequential recommendation representation learning, serving as an effective regularizer and optimizing the parameter space for the recommendation task.

\end{abstract}

%%
%% The code below is generated by the tool at http://dl.acm.org/ccs.cfm.
%% Please copy and paste the code instead of the example below.
%%
\begin{CCSXML}
<ccs2012>
 <concept>
  <concept_id>10010520.10010553.10010562</concept_id>
  <concept_desc>Computer systems organization~Embedded systems</concept_desc>
  <concept_significance>500</concept_significance>
 </concept>
 <concept>
  <concept_id>10010520.10010575.10010755</concept_id>
  <concept_desc>Computer systems organization~Redundancy</concept_desc>
  <concept_significance>300</concept_significance>
 </concept>
 <concept>
  <concept_id>10010520.10010553.10010554</concept_id>
  <concept_desc>Computer systems organization~Robotics</concept_desc>
  <concept_significance>100</concept_significance>
 </concept>
 <concept>
  <concept_id>10003033.10003083.10003095</concept_id>
  <concept_desc>Networks~Network reliability</concept_desc>
  <concept_significance>100</concept_significance>
 </concept>
</ccs2012>
\end{CCSXML}

% \ccsdesc[500]{Computer systems organization~Embedded systems}
% \ccsdesc[300]{Computer systems organization~Redundancy}
% \ccsdesc{Computer systems organization~Robotics}
% \ccsdesc[100]{Networks~Network reliability}
\ccsdesc[500]{Information systems~Recommender systems}

%%
%% Keywords. The author(s) should pick words that accurately describe
%% the work being presented. Separate the keywords with commas.
% \keywords{datasets, neural networks, gaze detection, text tagging}
\keywords{Multimodal Recommendation, Sequential Recommendation, Contrastive Learning}

% \received{20 February 2007}
% \received[revised]{12 March 2009}
% \received[accepted]{5 June 2009}

%%
%% This command processes the author and affiliation and title
%% information and builds the first part of the formatted document.
\maketitle

%%%%%%%%%%%%%%%%%%%%%%%%%%%%%%%%%%%%%%%%%%%%%%%%%%%%%%%%%%%%%%%%%%%%%%%%
\section{Introduction}
Sequential recommendation systems aim to capture users' dynamic preferences based on their historical behaviors, with the objective of predicting the next item of interest~\cite{wang2019sequential}.
The primary supervision signal for learning the parameters of these models typically derives from users' sequential interactions with items. However, \textcolor{black}{the sparsity of user behavior data poses significant challenges to sequential recommendation methods that solely rely upon it, often leading to suboptimal performance due to data sparsity~\cite{zhang2019feature,yuan2023go,wang2022transrec,pan2022multimodal,zhang2022diffusion}.} 
In practice, an abundance of multimodal content (\ie images and text descriptions) associated with items is available, \textcolor{black}{which} has been employed to mitigate the data sparsity issue in constructing traditional recommendation systems. 
For example, certain research~\cite{he2016vbpr,xu2018graphcar} 
% leverage item multimodal content as a regularization factor and integrate it with collaborative filtering frameworks. 
has leveraged item multimodal content as a regularization factor and incorporated it into collaborative filtering frameworks.
More recent research~\cite{wei2019mmgcn,wei2020graph,zhang2021mining} has deployed graph neural networks to discover hidden links between different modalities, 
thereby deepening the understanding of users' preferences.
\textcolor{black}{It has been demonstrated that multimodal data can substantially enhance the performance of recommendation systems.}

\textcolor{black}{Despite the considerable progress achieved by these multimodal methods, their application falls short in sequential recommendation as they are not explicitly designed to model the temporal dynamics of users' preferences for multimodal information.}
Currently, only a handful of studies exploit multimodal data for sequential recommendation. 
For instance, MV-RNN~\cite{cui2018mv} merges multimodal features at its input and employs a recurrent structure to dynamically track users' interests.
MML~\cite{pan2022multimodal} implements a modality-specific meta-learner to identify the sequential pattern from different modalities and adaptively merges their predictions using a learnable fusion layer.
Nevertheless, \textcolor{black}{existing multimodal sequential recommendation methods remain deficient in effectively exploring and capturing correlations among behavior sequences of users and items across different modalities.} Specifically,
1) existing works usually perform modality fusion (\ie concatenation, addition, or attention) at the item level, \textcolor{black}{thereby overlooking} the correlation among sequence representations in different modalities and the correlation between sequence representations with items in different modality spaces;  
2) existing works predominantly depend on a supervised learning framework that utilizes item prediction loss (\ie cross-entropy loss) to learn the entire model. This approach tends to overemphasize final performance while 
insufficiently capturing the association or fusion between multimodal data and sequence data.

In this work, we explore multimodal pre-training~\cite{bao2022vlmo,zhou2020unified,lu2019vilbert,wang2022image,radford2021learning} in the context of sequential recommendation with \textcolor{black}{the aim of enhancing multimodality fusion and utilization of multimodal information. The core idea of multimodal pre-training involves leveraging self-supervised signals to aggregate and align visual and textual information. Pre-training enables efficient exploitation of unlabeled data space, captures intrinsic data correlations, and ultimately improves the performance of downstream tasks. 
Although prevalent in computer vision, multimodal pre-training is relatively underexplored in sequential recommendation. Our objective is to integrate the advantages of multimodal pre-training into sequential recommendation representation learning. This is a non-trivial work as pre-training for sequential recommendation fundamentally differs from that for computer vision tasks. Multimodal pre-training for sequential recommendation predominantly focuses on modeling users' evolving preferences for multimodal information over time, rather than aligning images and text.}

To enhance multimodality fusion and more effectively harness multimodal information in sequential recommendation, 
\textcolor{black}{we propose a pre-training approach, \underline{M}ultimodal \underline{P}re-training for \underline{S}equential \underline{R}ecommendation (MP4SR). MP4SR utilizes contrastive losses to capture the correlation among behavior sequences of users and items across
different modalities.}
It comprises three key components: 
multimodal feature extraction, 
a backbone network Multimodal Mixup Sequence Encoder (M$^2$SE), 
and pre-training tasks.
Initially, we tokenize each item image into multiple text keywords using a language-image pre-trained model~\cite{radford2021learning}, and then apply the Sentence-BERT model~\cite{reimers-2019-sentence-bert} to extract initial text and image features of items.
This step eliminates discrepancies between the textual and visual modalities while preserving meaningful information from images and discarding redundant information. 
Next, M$^2$SE integrates different user modality sequences via a complementary sequence mixup strategy, subsequently processed by a Transformer to obtain mix-modality sequence representations.
Lastly, we 
employ contrastive learning to identify modality interactions at both sequence-to-sequence and sequence-to-item levels.
Specifically, we use a \textit{modality-specific next item prediction loss} to capture the correlation between a mix-modality sequence and the subsequent item within each modality space, and a \textit{cross-modality contrastive learning loss} to calibrate the discrepancies of mix-modality sequence representations across different modality spaces.

We conduct extensive experiments on four real-world datasets to evaluate the effectiveness of MP4SR. Our experimental results show that MP4SR outperforms state-of-the-art approaches for sequential recommendation under both normal and cold-start settings. We also show that restricting the feasible starting points in parameter space to those minimizing the self-supervised pre-training criterion yields similar effects to a good regularizer on the parameters, thereby enhancing the recommendation performance.

%%%%%%%%%%%%%%%%%%%%%%%%%%%%%%%%%%%%%%%%%%%%%%%%%%%%%%%%%%%%%%%%%%%%%%%%

\section{Related Work}

\subsection{Sequential Recommendation}
Earlier works~\cite{rendle2010factorizing} on sequential recommendation adopt Markov Chain to capture the transitions over user-item interaction sequences. However, 
these methods are not well suited to handle complex sequence patterns.
More recently, deep learning techniques such as Convolutional Neural Networks~\cite{tang2018personalized,yuan2019simple}, Recurrent Neural Networks~\cite{donkers2017sequential,peng2021ham}, and Graph Neural Networks~\cite{chang2021sequential,IJCAI-GCL4SR} have been applied to model users' sequential behaviors.
Moreover, transformer architecture-based methods~\cite{kang2018self,sun2019bert4rec,zhou2020s3,liu2021augmenting} have shown strong performance in capturing long-range dependencies in a sequence. 

\textcolor{black}{To improve the performance of the sequential recommendation, 
some recent studies integrate auxiliary information (\ie item descriptions~\cite{hou2022towards,zhang2019feature,zhang2024greenrec}, item images~\cite{wei2020graph, wang2021dualgnn}, user profiles~\cite{wu2022personalized,xu2018graphcar}, review texts~\cite{wang2021leveraging,shuai2022review}, etc) into the sequential recommendation framework.}
For example, in FDSA~\cite{zhang2019feature}, different item features are first aggregated using a vanilla attention layer, followed by a feature-based self-attention block to learn how features transit among items in a sequence. The $\textrm{S}^3$-Rec~\cite{zhou2020s3} adopts a pre-training strategy to predict the correlation between an item and its attributes. 
Moreover, in DIF-SR~\cite{xie2022decoupled}, the auxiliary information is moved from the input to the attention layer. The attention calculation of auxiliary information and item representations is decoupled to improve the modeling capability of item representations. 
UniSRec~\cite{hou2022towards} leverages item text representations with an MoE-based adaptor and employs contrastive learning tasks to learn transferable sequence representations.
While these methods have utilized auxiliary information across various modalities, the majority tend to rely on features extracted from a singular modality and have yet to explore the integration of information derived from multiple modalities.
In this work, our focus is on the fusion of multimodal information of items with user sequential behaviors to enhance the effectiveness of sequential recommendations.

\subsection{Multimodal Recommendation}
Multimodal recommendation methods exploit the multimodal content of items to improve recommendation performance. Previous works~\cite{he2016vbpr,kang2017visually,xu2018graphcar} incorporate the visual features of item images into the matrix factorization-based recommendation framework. 
In other works~\cite{liu2019user,liu2019userdiverse,liu2021pre}, attention networks are used to combine multimodal features to enhance the representation learning of users and items. For example, UVCAN~\cite{liu2019user} learns multi-modal information about users and items using stacked attention networks. \textcolor{black}{Moreover, several recent studies~\cite{wei2019mmgcn,wei2020graph,zhang2021mining,wang2021dualgnn,yi2021multi,BM3WWW2023,FREEDOMMM2023,zhou2023enhancing} leverage GNNs to exploit the item's multimodal information.}
For example, LATTICE~\cite{zhang2021mining} embeds a modality-aware graph structure learning layer that identifies item-item graph structures using multimodal features.
This layer injects high-order item affinities into item representations explicitly. \textcolor{black}{On top of LATTICE, FREEDOM~\cite{FREEDOMMM2023} freezes the item-item graph to achieve a recommendation that is both effective and efficient.}
DualGNN~\cite{wang2021dualgnn} explicitly models the user's attention over different modalities and inductively learns her multi-modal preference. In MVGAE~\cite{yi2021multi}, a multi-modal variational graph auto-encoder is proposed to fuse modality-specific node embeddings according to the product-of-experts principle.

\textcolor{black}{Several studies~\cite{cui2018mv,pan2022multimodal,wang2022transrec,liang2023mmmlp,hu2023adaptive,zhang2024dual,zhang2024id,zhou2023comprehensive,lei2021semi} have utilized multimodal information for the sequential recommendation.} 
For instance, MV-RNN~\cite{cui2018mv} proposes to combine item latent embedding with multimodal features using concatenation, addition, or feature reconstruction. The user's sequential behaviors are captured by a recurrent neural network. 
\textcolor{black}{TransRec~\cite{wang2022transrec} investigates pre-trained recommender systems through an end-to-end approach, mixing up item features across different modalities.}
\textcolor{black}{Inspired by the success of MLP-Mixer~\cite{tolstikhin2021mlp} in CV, MMMLP~\cite{liang2023mmmlp} adapts the architecture to multimodal sequential recommendation and is a purely MLP-based architecture to have an edge on both efficacy and efficiency.
However, these approaches have yet to explicitly and effectively capture the intricate correlations existing between user behavior sequences and the multimodal information of items. To address this issue, we propose a pre-training and fine-tuning paradigm, which employs contrastive learning with a complementary sequence mixup strategy. 
}

\subsection{Multimodal Pre-training}
Recently, a significant number of multimodal pre-trained models~\cite{bao2022vlmo,zhou2020unified,lu2019vilbert,wang2022image,radford2021learning} have been developed to learn generic cross-modal representations from large-scale image-text pairs in the computer vision community. These vision language pre-training models can be categorized into dual-encoder, encoder-decoder, and fusion-encoder architecture. 
Dual-encoder methods~\cite{radford2021learning,jia2021scaling} use separate encoders to embed the language and image data into embeddings. The cosine similarity between image and text feature vectors is used in order to account for modality interaction.
Encoder-decoder methods~\cite{wang2021simvlm} recover corrupted input tokens for the generation task.
Fusion-encoder methods~\cite{lu2019vilbert,su2019vl,kim2021vilt} leverage the cross-modal attention to model image-text pairs. 
\textcolor{black}{In this paper, we present a novel framework that harnesses multimodal pre-training for sequential recommendation. Our approach to multimodal pre-training differs from the existing research, which primarily focuses on aligning images and text. In the context of sequential recommendation, multimodal pre-training need to deal with the modelling of users’ evolving preferences for multimodal information of items over time.}

%%%%%%%%%%%%%%%%%%%%%%%%%%%%%%%%%%%%%%%%%%%%%%%%%%%%%%%%%%%%%%%%%%%%%%%%
\begin{figure*}[t]
\centering
\includegraphics[width=\textwidth]{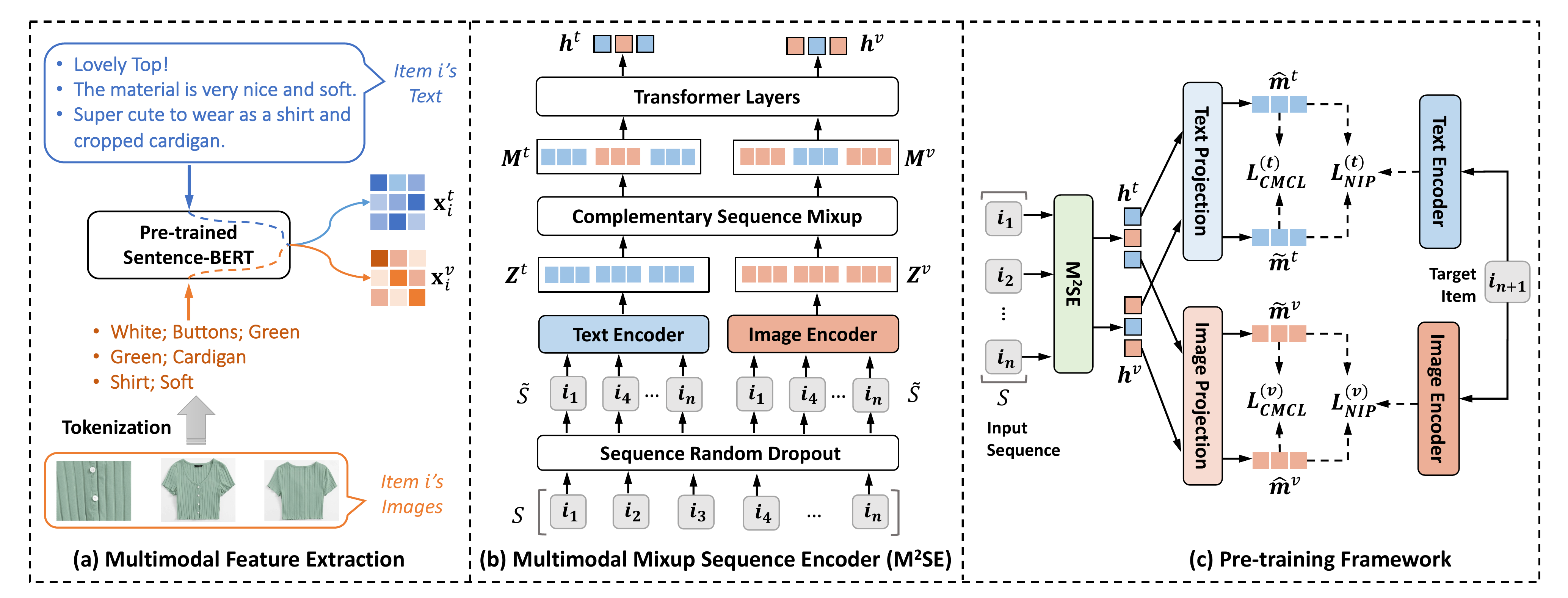}
\caption{Overall framework of the proposed method MP4SR, which consists of three main components: (a) The multimodal feature extraction module used to obtain initial multimodal features of items. (b) The structure of the proposed multimodal mixup sequence encoder that fuses items' multimodal content with users' behavior sequence. (c) The workflow of the proposed pre-training framework, where $\C{S}$ is the input sequence and $i_{n+1}$ is the target item. }
\label{fig:framework}
\end{figure*}

\section{Methodology}
In this section, we introduce the main components of MP4SR, including multimodal feature extraction, backbone network M$^2$SE, pre-training objectives, and fine-tuning objectives.

\subsection{Notations}

Let $\mathcal{I}$ denote the set of items, and $\C{S} = \{i_1, i_2, \cdots, i_n\}$ denote a user behavior sequence, where $n$ items are sorted in chronological order based on the interaction timestamp. In this work, we consider the text and image content of items to build the model. For each item $i$, it is associated with a chunk of text descriptions that is split into sentences as $\C{T}_i=\{t_1^i, t_2^i, \cdots , t^i_{|\mathcal{T}_i|}\}$, and a set of images $\C{V}_i=\{v_1^i, v^i_2, \cdots, v^i_{|\C{V}_i|}\}$, where $|\C{T}_i|$ and $|\C{V}_i|$ denote the number of sentences and images, respectively.

\subsection{Multimodal Feature Extraction}
\textcolor{black}{Figure~\ref{fig:framework}(a) shows the workflow using pre-trained models to obtain the initial text and image features of items to eliminate the modality gap between the text and image embeddings.} 

\subsubsection{Text Feature Extraction}
For each sentence in $\C{T}_i$, we feed it into the pre-trained Sentence-BERT~\cite{reimers-2019-sentence-bert} to obtain its latent representation. The initial text feature $\B{x}_i^t$ of item $i$ is obtained by stacking representations of all the sentences in $\C{T}_i$ as follows,
\begin{equation}
    \B{x}_i^t = stack\big[\textrm{BERT}(t^i_1), \textrm{BERT}(t^i_2), \cdots, \textrm{BERT}(t^i_{|\mathcal{T}_i|})\big],
\end{equation}
where $\B{x}_i^t\in\mathbb{R}^{|\mathcal{T}_i|\times d}$, $stack[,]$ denotes stacking multiple vectors into a matrix, and $d$ is the embedding dimension.

\subsubsection{Image Feature Extraction}~\label{sec:clip} \textcolor{black}{Inspired by~\cite{lin2022towards}, we use a pre-trained language-image model, \ie CLIP~\cite{radford2021learning}, to describe each image by text tokens in order to eliminate the modality gap between text and image representations and remove irrelevant information from images.} 
To capture the key visual information of an image, $N$ most relevant text tokens are retained based on their similarities to the image. Then, we obtain the initial feature $\B{v}^i_\ell$ for an item image $v^i_\ell \in \C{V}_i$, by concatenating these text tokens as a sentence and feeding it into the same pre-trained Sentence-BERT model. The initial image feature $\B{x}_i^v$ of item $i$ can be obtained by stacking the features of all images in $\C{V}_i$ as follows,
\begin{align}
    f(w) &= \textrm{sim}(\textrm{CLIP}(v^i_\ell), \textrm{CLIP}(w)) \qquad \forall w \in \C{D}, \nonumber\\
    \B{v}^i_{\ell}&=\textrm{BERT}(concat(\textrm{TopN}(\{f(w_1),\cdots,f(w_{|\mathcal{D}|})\}, N))), \nonumber\\
    \B{x}_i^v &= stack[\B{v}^i_{1}, \B{v}^i_{2}, \cdots, \B{v}^i_{|\mathcal{V}_i|}],
\label{eq:image2text}
\end{align}
where $\B{x}_i^v \in \mathbb{R}^{|\C{V}_i|\times d}$, $w$ is a text token in the word dictionary $\C{D}$, and $|\mathcal{D}|$ denotes the size of the dictionary. $\textrm{sim}(\cdot)$ is to compute the cosine similarity between the embedding of an image $v_\ell^i$ and a word $w$ obtained by the CLIP model.
$\textrm{TopN}(\cdot)$ function selects $N$ words that have the highest similarities with the image. $concat(\cdot)$ is the operation of concatenating $N$ words into one sentence. Note that $\B{x}_i^t$ and $\B{x}_i^v$ are derived during the data pre-processing stage.

\begin{figure*}[!t]
\centering
\subfloat[Pantry]{\includegraphics[width=0.48\textwidth]{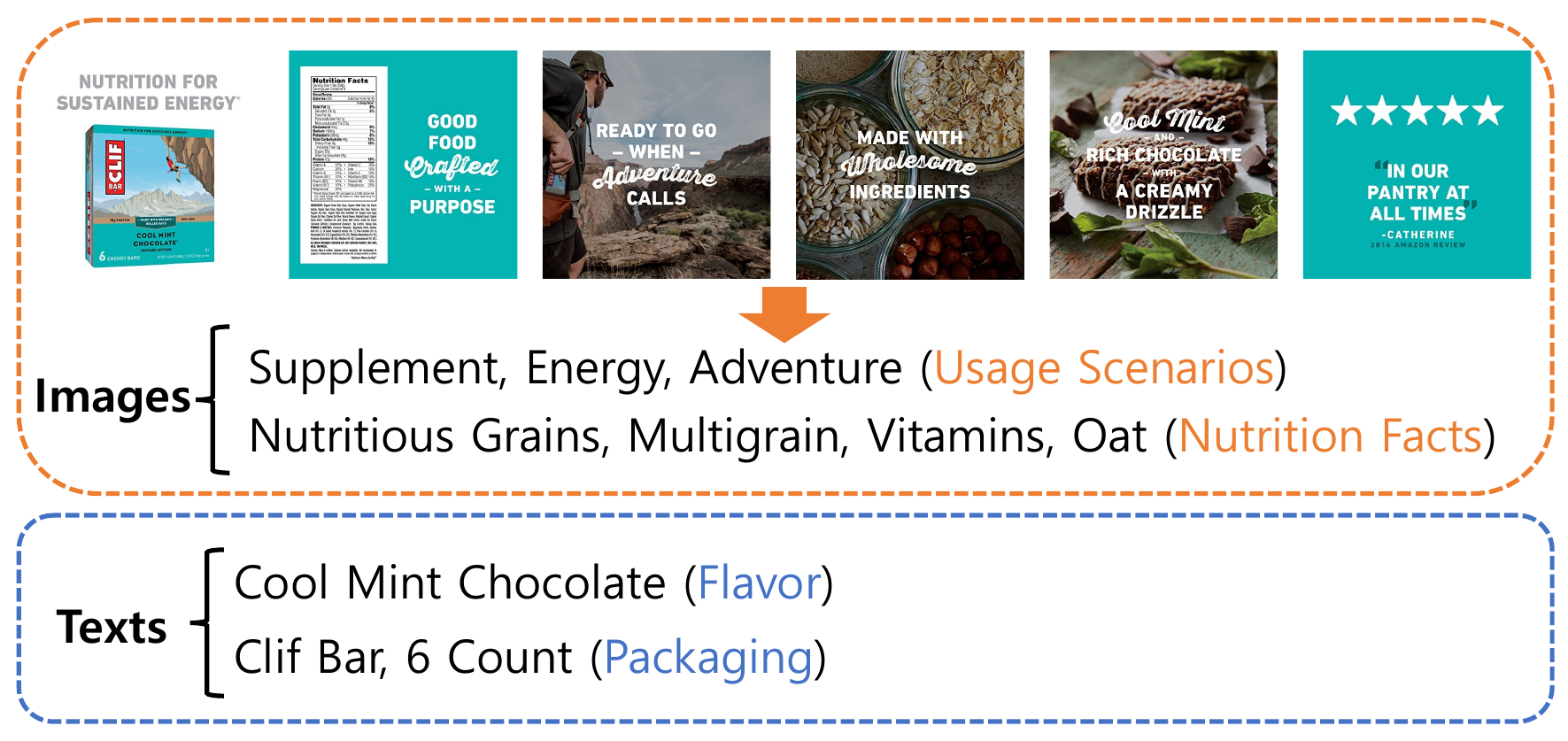}}
\quad
\subfloat[Arts]{\includegraphics[width=0.48\textwidth]{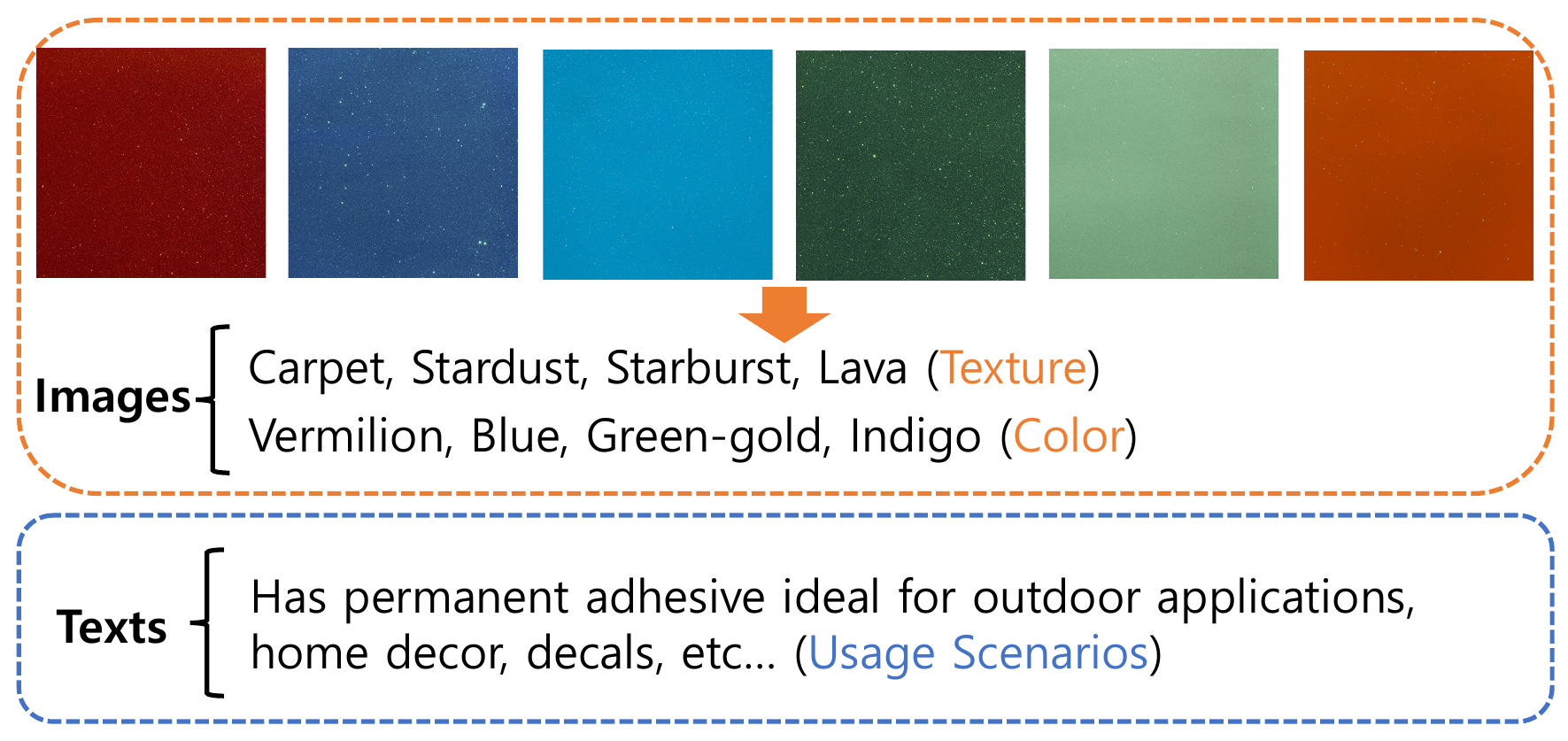}}
\caption{Two examples of converting images of an item into text tokens. Items are retrieved from the Amazon Pantry and Arts dataset. Text tokens are generated using CLIP~\cite{radford2021learning}.}
\label{fig:image-tokens}
\end{figure*}

\textcolor{black}{To verify text tokens retrieved from images using the pre-trained language-image model, we select two representative items from the Amazon Pantry and Arts dataset for analysis~(Figure~\ref{fig:image-tokens}). 
The left item is an energy bar to provide carbohydrates and protein, with incomplete information in the seller's summary. 
By generating text tokens from images, key information, including nutritional facts and usage modes, can be obtained. 
The right item is a carpet with multiple colors, and the merchant's product descriptions lack information on the carpet's appearance. 
By using text tokens from images, we can describe the characteristics of the carpet based on its colors and designs.
Compared to the image feature extracted from pre-trained image encoders (\eg ResNet~\cite{wei2019mmgcn,zhang2021mining}), word tokens converted from images can discard irrelevant information to achieve better recommendation performance. }

\subsection{Multimodal Mixup Sequence Encoder}

\textcolor{black}{To encode user sequences with multimodal features extracted from items, we explain the backbone network, \ie M$^2$SE.} The structure of M$^2$SE is shown in Figure~\ref{fig:framework}(b). Observe that M$^2$SE includes four main components: sequence random dropout, text and image encoders, complementary sequence mixup, and transformer layers. 

\subsubsection{Sequence Random Dropout}
\textcolor{black}{To help the model achieve better generalization performance, M$^2$SE randomly drops a portion of items from $\C{S}$ with a drop ratio $\rho$ for a user behavior sequence $\C{S}$~\cite{cheng2021learning,xie2022contrastive}.} The obtained sub-sequence after the random dropout operation is denoted by $\widetilde{\C{S}}$.
\textcolor{black}{$\rho$ is fixed during the pre-training stage.}

\subsubsection{Text and Image Encoders}
These two encoders are used to adapt the initial modality features of items obtained from the pre-trained language model to learn users' sequential behaviors. Both encoders share the same structure, including an attention layer and a Mixture-of-Expert (MoE) architecture~\cite{shazeer2017outrageously}.

In the text encoder, each item $i \in \widetilde{\C{S}}$ is represented by its initial textual feature $\B{x}_i^t$. 
The attention layer is composed of two linear transformations to fuse $i$'s sentence-level embeddings as follows,
\begin{align}
    \B{\alpha}^t &= \textrm{softmax}\big((\B{x}_i^t\B{W}^{t}_{1} + \B{b}^{t}_{1})\B{W}^{t}_{2}+b^{t}_{2}\big), \nonumber\\
    \B{e}_i^t &= \sum_{j=1}^{|\C{T}_i|}\alpha^t_j\B{x}_i^t[j,:],
\end{align}
where $\B{W}^{t}_{1}\in\mathbb{R}^{d\times d_a}$, $\B{W}^{t}_{2}\in\mathbb{R}^{d_a}$, $\B{b}^{t}_{1}\in\mathbb{R}^{d_a}$, and $b^{t}_{2}\in\mathbb{R}$ are learnable parameters. $d_a$ is the attention dimension size. $\alpha^t_j$ is the $j$-th element of $\alpha^t$, and $\B{x}_i^t[j,:]$ denotes the $j$-th row of feature matrix $\B{x}_i^t$. Then, MoE is used to increase the model’s capacity for adapting the fused modality representation $\B{e}_i^t$. Each expert in MoE consists of a linear transformation, followed with a dropout layer and a normalization layer. Let $E_k(\B{e}_i^t)\in\mathbb{R}^{d_0}$ denote the output of the $k$-th expert network, and $\B{g}^t\in\mathbb{R}^{O}$ is the output of the gating network as follows,
\begin{align}
    E_k(\B{e}_i^t) &= \textrm{LayerNorm}(\textrm{Dropout}(\B{e}_i^t\B{W}_k^t)), \nonumber\\
    \textbf{g}^t &= \textrm{softmax}(\B{e}_i^t\B{W}^t_{3}),
\end{align}
where $\B{W}^{t}_{3}\in\mathbb{R}^{d\times O}$ and $\textbf{W}_k^t\in\mathbb{R}^{d\times d_0}$ are learnable parameters, $O$ is the number of experts, and $d_0$ is the dimension of the hidden embedding. Then, the output of MoE for item $i$ is formulated as follows,
\begin{equation}
    \B{z}_i^t = \sum_{k=1}^{O}g_k^t E_{k}(\B{e}_i^t),
    \label{eq:x}
\end{equation}
where $\textbf{z}_i^t \in \mathbb{R}^{d_0}$, and $g^t_k$ is the weight derived from $k$-th gating router. Here, we omit bias terms in the equation for simplicity.
The outputs of MoE network for all items in $\widetilde{\C{S}}$ are stacked to form the output of the text encoder, which is denoted by $\B{Z}^t=stack[\B{z}_1^t, \B{z}_2^t, \cdots, \B{z}_{|\widetilde{\C{S}}|}^t]$.

Similarly, in the image encoder, each item in $\widetilde{\C{S}}$ is represented by its image feature $\B{x}_i^v$. The output of the image encoder is denoted by $\B{Z}^v=stack[\B{z}_1^v, \B{z}_2^v, \cdots, \B{z}_{|\widetilde{\C{S}}|}^v]$, where $\B{z}_i^v$ is the output of the MoE network for the $i$-th item in $\widetilde{\C{S}}$.

\subsubsection{Complementary Sequence Mixup}

To alleviate the representation discrepancy between two different modality sequences, we propose a complementary sequence mixup method that mixes up text representations and image representations in a complementary manner. Specifically, we define a mixup ratio $p$ between $0$ to $0.5$, which is randomly generated during model training. For each item in $\widetilde{\C{S}}$, we swap its embedding in $\B{Z}^t$ and $\B{Z}^v$ with probability $p$ and generate two mix-modality sequence embeddings $\B{M}^t$ and $\B{M}^v$. The definition of $p \le 0.5$ ensures the generated mix-modality sequence embedding dominates with information from the same modality. In this case, $\B{M}^t$ and $\B{M}^v$ complement each other in terms of the modality choice for each item in the sequence.

\subsubsection{Transformer Layers}

The Transformer~\cite{vaswani2017attention} structure is used to further encode $\B{M}^t$ and $\B{M}^v$.
We first add positional encodings to $\B{M}^t$ and $\B{M}^v$, and then feed the summed embeddings into $L$ Transformer layers. Note that each Transformer layer consists of a multi-head self-attention sub-layer and a point-wise feed-forward network. Let $\B{H}^t_{L}$ and $\B{H}^v_{L}$ denote the output of the $L$-th Transformer layer based on $\B{M}^t$ and $\B{M}^v$, respectively.
Following~\cite{zhou2020s3}, we use the last rows in $\B{H}^t_{L}$ and $\B{H}^v_{L}$ as two mix-modality representations of the input sequence, which are denoted by $\B{h}^t$ and $\B{h}^v$.

\subsection{Pre-training Objectives}
\textcolor{black}{To better capture the correlation between representations across different modalities, we propose two optimization objectives, \ie modality-specific next item prediction and cross-modality contrastive learning, based on mix-modality sequence representations for pre-training the backbone model.}
The workflow in the pre-training phase is shown in Figure~\ref{fig:framework}(c). 

Let $\C{B}=\{(\C{S}_j, i_j)\}_{j=1}^{|\C{B}|}$ denote a batch of pre-training data, where $\C{S}_j$ denotes a user's behavior sequence and $i_j$ is her next interaction item after $\C{S}_j$. With M$^2$SE, we can obtain two mix-modality sequence representations $\B{h}_j^{t}$ and $\B{h}_j^{v}$ for $\C{S}_j$. As $\B{h}_j^t$ and $\B{h}_j^v$ are obtained by mixing up modalities, we first use two linear transformations to map them into the text feature space and image feature space \textcolor{black}{for calculating the pre-training losses}, respectively,
\begin{align}
    \widehat{\B{m}}_j^{t} &= \B{h}^{t}_j\B{W}_t + \B{b}_t,
    & \widetilde{\B{m}}_j^{t} &= \B{h}_j^{v}\B{W}_t + \B{b}_t, \nonumber\\
    \widehat{\B{m}}_j^{v} &= \B{h}_j^{v}\B{W}_v + \B{b}_v,
    & \widetilde{\B{m}}_j^{v} &= \B{h}_j^{t}\B{W}_v + \B{b}_v,
\end{align}
where $\B{W}_t, \B{W}_v\in\mathbb{R}^{d_0\times d_0}$ and $\B{b}_t,\B{b}_v\in\mathbb{R}^{d_0}$ are learnable parameters.
Motivated by the success of contrastive learning in model pre-training, we define the pre-training objective functions in a contrastive manner.

\subsubsection{Modality-specific Next Item Prediction}
Modality-specific Next Item Prediction (NIP) aims to predict the next item based on the mix-modality sequence representations.
For each $(\C{S}_j, i_j)$ pair, $\C{S}_j$ is the input sequence and $i_j$ is the target item. Thus, in the text feature space, we pair $\widehat{\B{m}}_j^t$ and $\widetilde{\B{m}}_j^t$ with the $i_j$'s text embedding $\B{z}_j^t$ obtained by the text encoder as a positive sample, and pair $\widehat{\B{m}}_j^t$ and $\widetilde{\B{m}}_j^t$ with the text embeddings of other items $\{i_{j'}|j' \neq j, 1 \leq j' \leq |\C{B}|\}$ from $\C{B}$ as negative samples. The next item prediction loss defined in the text feature space is as follows,
\begin{align}
    \mathcal{L}_\textrm{NIP}^{(t)} =
    -\sum^{|\C{B}|}_{j=1} \log\frac{
    f(\widehat{\B{m}}_{j}^t, \B{z}^{t}_{j}) +
    f(\widetilde{\B{m}}_{j}^t, \B{z}^{t}_{j})}
    {\sum_{j'=1}^{|\C{B}|} [f(\widehat{\B{m}}_{j}^t, \B{z}^{t}_{j'}) + f(\widetilde{\B{m}}_{j}^t, \B{z}^{t}_{j'})]},
\end{align}
where $ f(\B{s}, \B{z}) = \textrm{exp}(\textrm{sim}(\B{s}, \B{z})/\tau)$, and $\tau$ is a temperature hyper-parameter. Similarly, we can define the next item prediction loss $\mathcal{L}_\textrm{NIP}^{(v)}$ in the image feature space as follows,
\begin{align}
    \mathcal{L}_\textrm{NIP}^{(v)} =
    -\sum^{|\C{B}|}_{j=1} \log\frac{
    f(\widehat{\B{m}}_{j}^v, \B{z}^{v}_{j}) +
    f(\widetilde{\B{m}}_{j}^v, \B{z}^{v}_{j})}
    {\sum_{j'=1}^{|\C{B}|} [f(\widehat{\B{m}}_{j}^v, \B{z}^{v}_{j'}) + f(\widetilde{\B{m}}_{j}^v, \B{z}^{v}_{j'})]}.
\end{align}

\subsubsection{Cross-Modality Contrastive Learning}
\textcolor{black}{To capture the semantic relationship between different modality sequences}, we develop a Cross-Modality Contrastive Loss (CMCL). Specifically, the complementary mix-modality sequence representations mapped to the same feature space, \eg ($\widehat{\B{m}}_j^t$, $\widetilde{\B{m}}_j^t$) and ($\widehat{\B{m}}_j^v$, $\widetilde{\B{m}}_j^v$), are paired as positive samples, while randomly-selected samples in the training batch are paired as negative samples. 
Following~\cite{zhu2021graph}, CMCL for the text space is defined in a symmetric contrastive way as follows,
\begin{align}
    &\ell(\widehat{\B{m}}_{j}^t, \widetilde{\B{m}}_{j}^t) =
    \log\frac{
    f(\widehat{\B{m}}_{j}^t,\widetilde{\B{m}}_{j}^t)}
    {\sum\limits_{j'=1}^{|\C{B}|} f(\widehat{\B{m}}_{j}^t, \widetilde{\B{m}}_{j'}^t) +
    \sum\limits_{j'=1,j'\neq j}^{|\C{B}|} f(\widehat{\B{m}}_{j}^t, \widehat{\B{m}}_{j'}^t)},\nonumber\\
    &\mathcal{L}_\textrm{CMCL}^{(t)} = -\frac{1}{2}\sum_{j=1}^{|\C{B}|}(
    \ell(\widehat{\B{m}}_{j}^t, \widetilde{\B{m}}_{j}^t) +
    \ell(\widetilde{\B{m}}_{j}^t, \widehat{\B{m}}_{j}^t)).
\end{align}
Similarly, CMCL in the image space is defined as follows,
\begin{equation}
    \mathcal{L}_\textrm{CMCL}^{(v)} = -\frac{1}{2}\sum_{j=1}^\C{B}(
    \ell(\widehat{\B{m}}_{j}^v, \widetilde{\B{m}}_{j}^v) +
    \ell(\widetilde{\B{m}}_{j}^v, \widehat{\B{m}}_{j}^v)).
\end{equation}

The overall loss function for model pre-training is formulated as follows,
\begin{equation}
    \mathcal{L}_{\textrm{pre-train}} = \mathcal{L}_\textrm{NIP}^{(t)} + \mathcal{L}_\textrm{NIP}^{(v)} + \lambda(\mathcal{L}_\textrm{CMCL}^{(t)} + \mathcal{L}_\textrm{CMCL}^{(v)}),
\end{equation}
where $\lambda$ is a hyper-parameter to balance these two groups of losses.

\begin{figure}[!t]
    \centering
    \subfloat[Pantry: Transformer $L$=1]{\includegraphics[width=0.3\columnwidth]{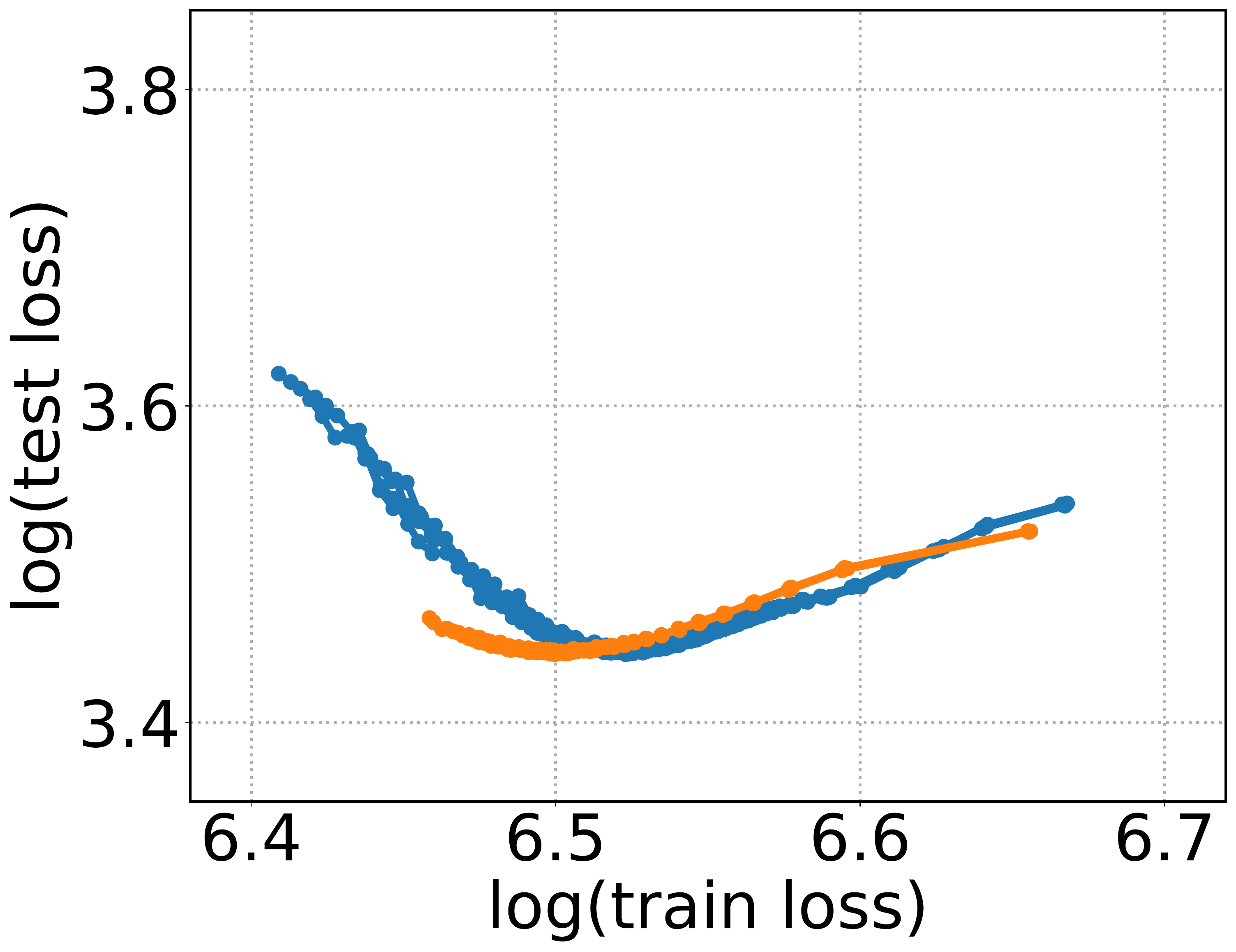}}
    \quad
    \subfloat[Pantry: Transformer $L$=2]{\includegraphics[width=0.3\columnwidth]{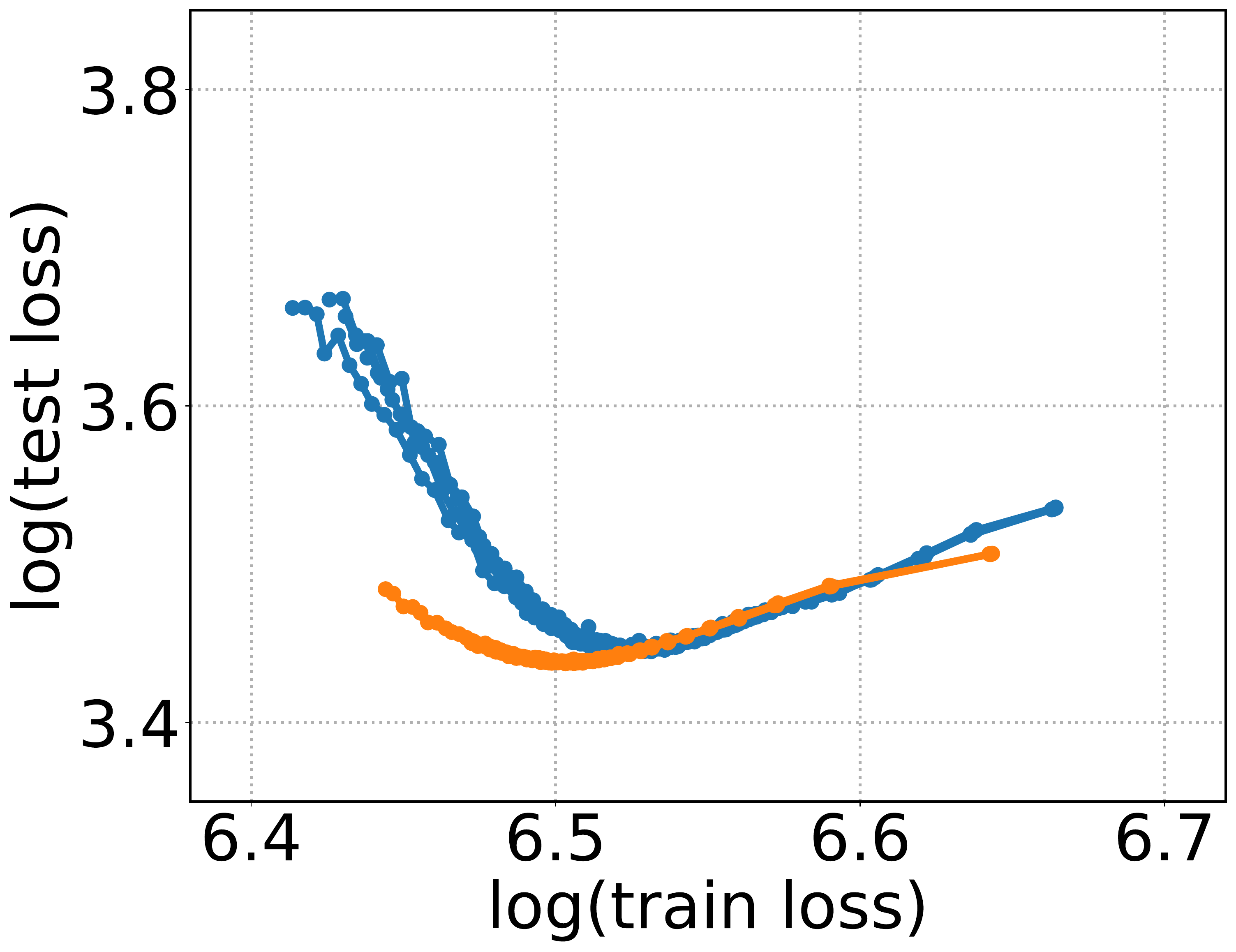}}
    \quad
    \subfloat[Pantry: Transformer $L$=3]{\includegraphics[width=0.3\columnwidth]{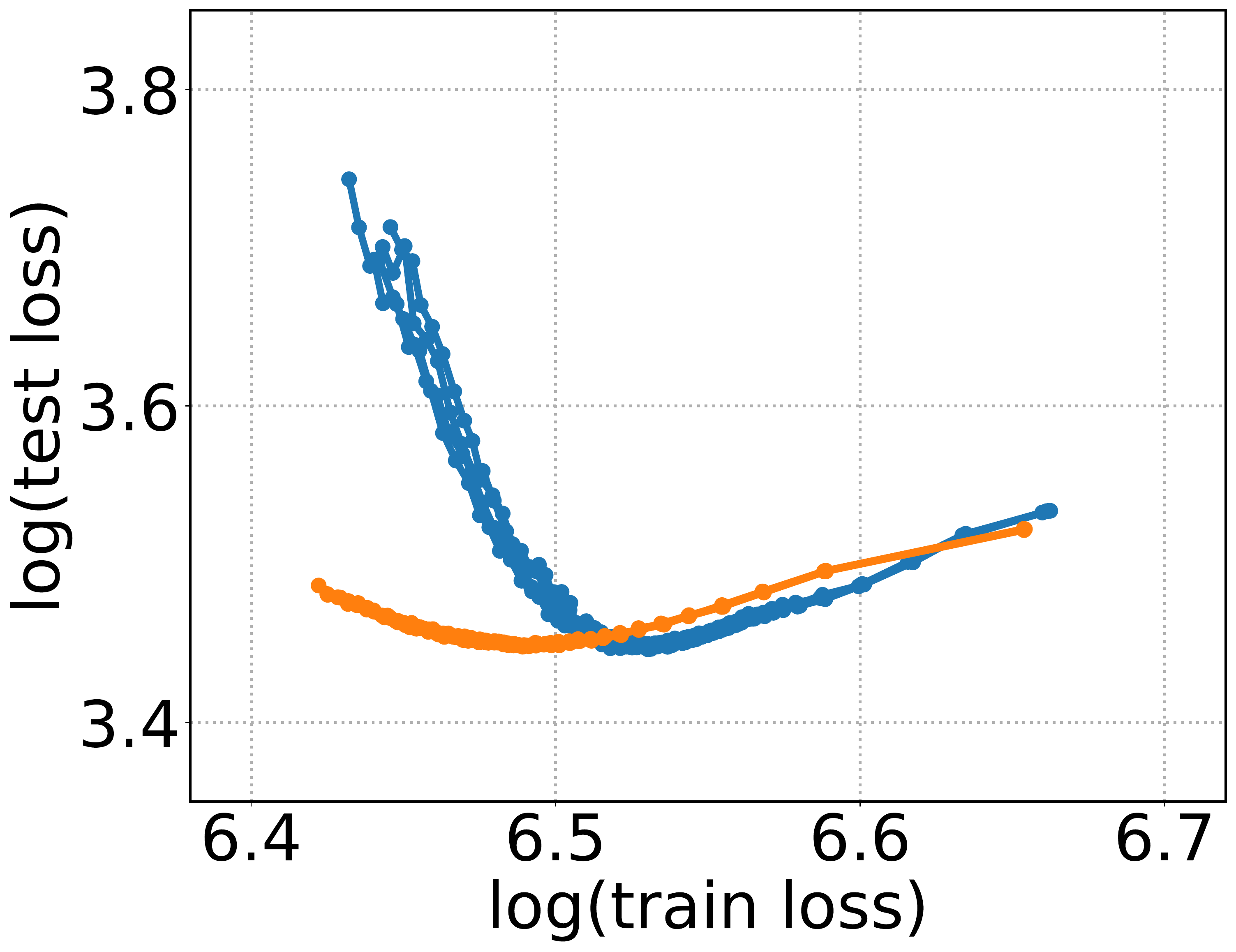}} \\
    \subfloat[Office: Transformer $L$=1]{\includegraphics[width=0.3\columnwidth]{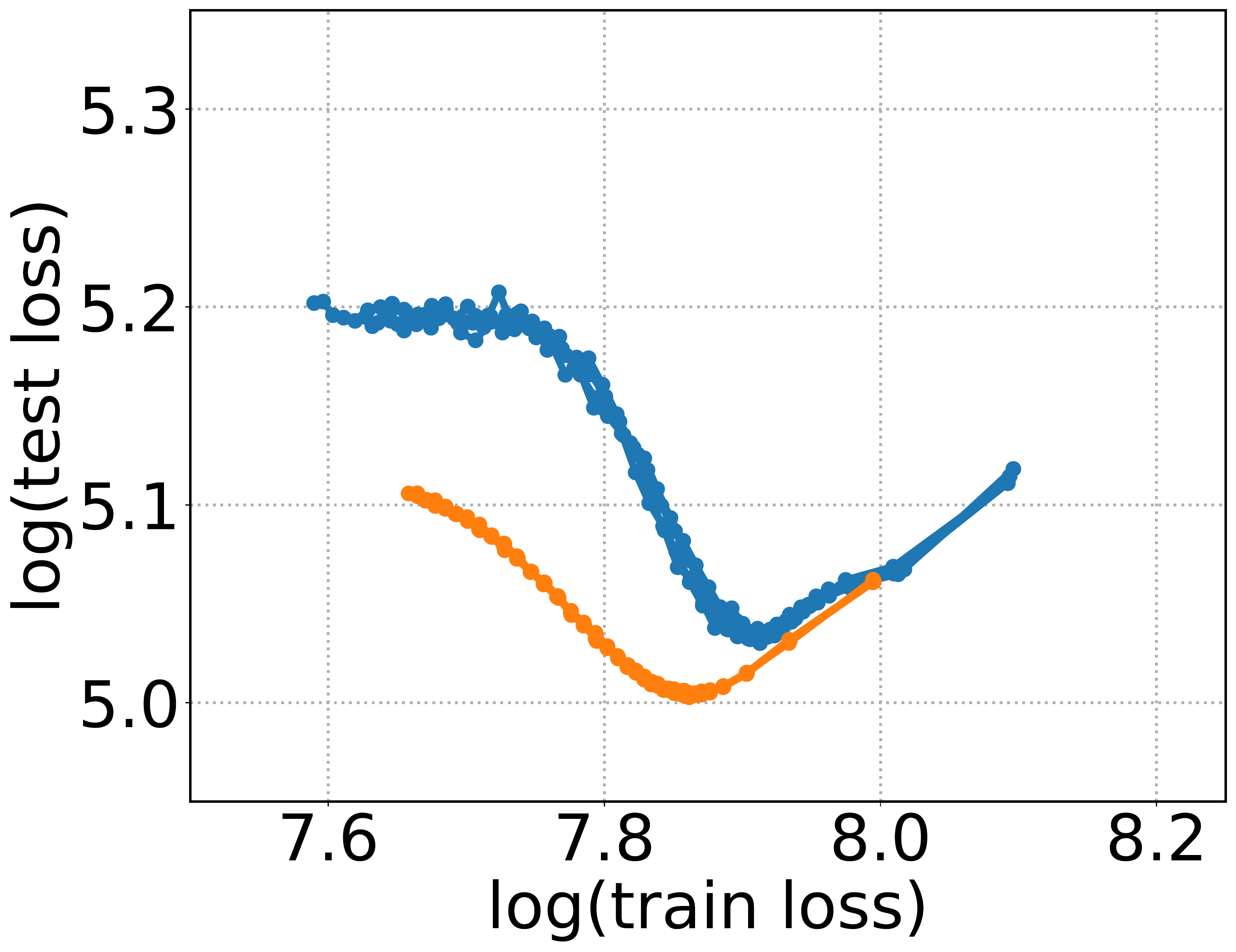}}
    \quad
    \subfloat[Office: Transformer $L$=2]{\includegraphics[width=0.3\columnwidth]{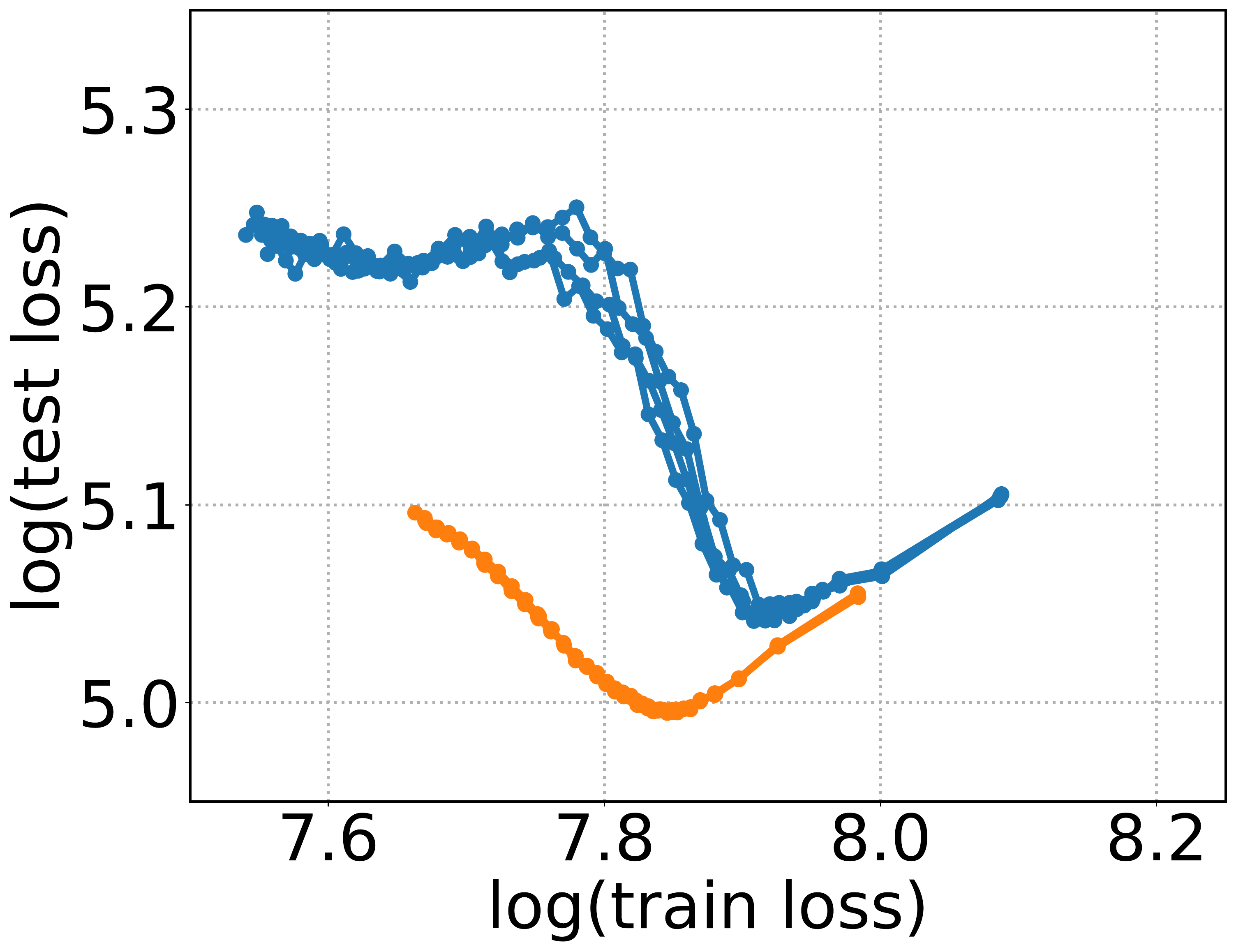}}
    \quad
    \subfloat[Office: Transformer $L$=3]{\includegraphics[width=0.3\columnwidth]{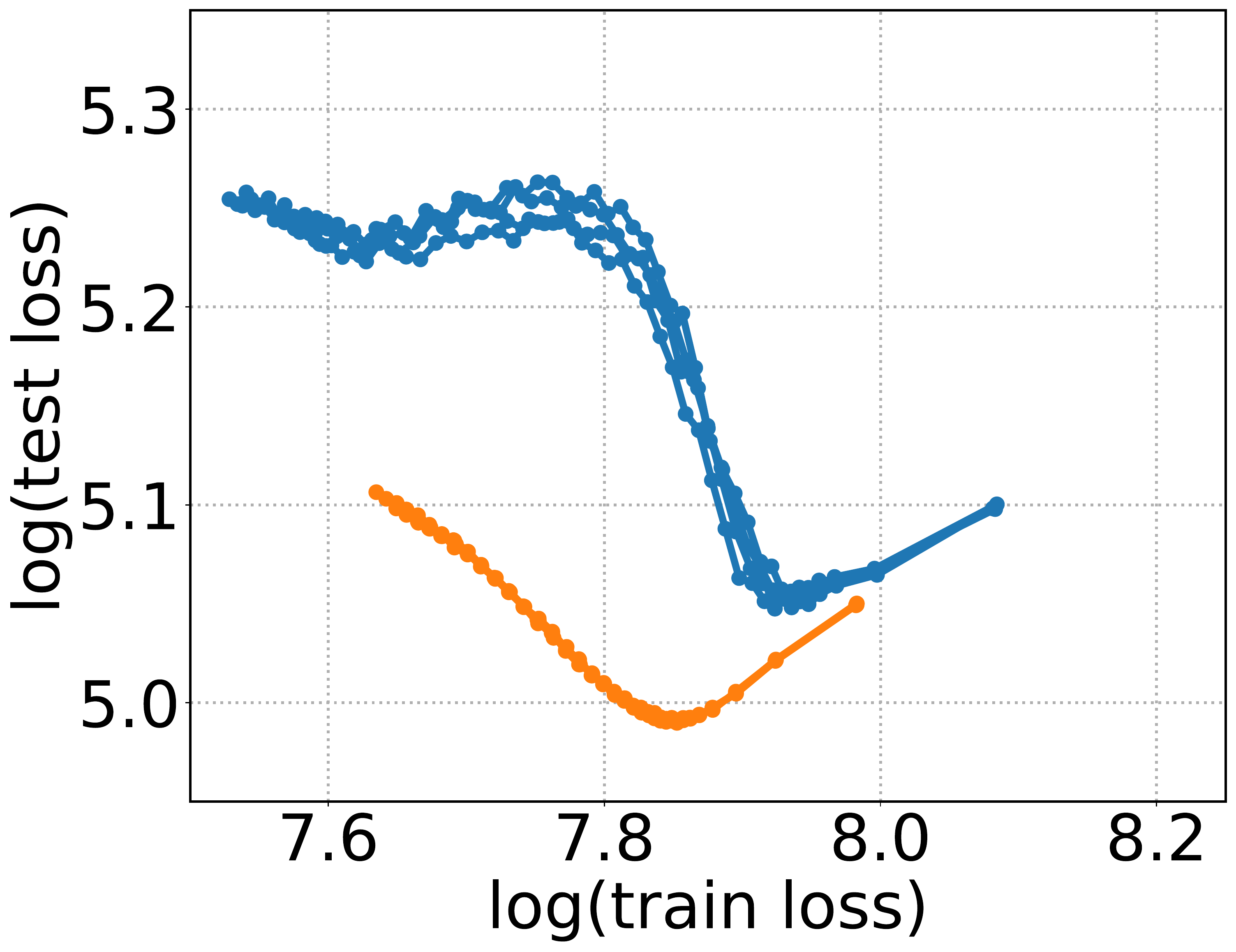}}
    \caption{Evolution without pre-training (\textcolor{ACMDarkBlue}{blue}) and with pre-training (\textcolor{orange}{orange}) on Pantry and Office datasets of the log of the test loss plotted against the log of the train loss as training proceeds. Each group has 5 curves representing a different initialization. During training, the trajectories move from right (high error) to left (low error) due to the decrease in training error.} 
\label{fig:loss}
\end{figure}

\subsubsection{Discussion}
\textcolor{black}{Inspired by~\cite{erhan2010does}, we plot test loss against train loss along the optimization trajectory in parameter space. 
Figure~\ref{fig:loss} shows five of these curves without pre-training (blue), originating from a random initialization point in parameter space, and five initiated from pre-trained parameters (orange).
The experiments are performed for MP4SR with 1, 2, and 3 Transformer layers.
We observe that, at the same level of train loss, pre-trained models consistently exhibit a lower test cost than randomly initialized models when training proceeds to convergence.
This finding suggests that pre-training tasks serve as effective regularizers for parameters used for modeling user sequences based on multimodal data.
Additionally, as the number of layers increases, the impact of generalization becomes more pronounced, which is evident from the downward shift of the orange line.}

\subsection{Fine-tuning for Sequential Recommendation}

\textcolor{black}{We treat sequential recommendation as a supervised classification problem~\cite{kang2018self,zhou2020s3} and use the cross-entropy loss incorporating all modality sequences for model fine-tuning.}

Let $\widetilde{\C{B}}$ denote a batch of fine-tuning data. For each $(\C{S}, i) \in \widetilde{\C{B}}$, $i$ is the user's next interaction item after her interaction sequence $\C{S}$. From the classification perspective, $i$ can be treated as the target label of the input sequence $\C{S}$.
In the fine-tuning stage, we disable the sequence random dropout and complementary sequence mixup operations in the pre-trained M$^2$SE network, by setting the dropout ratio $\rho$ and sequence mixup ratio $p$ to 0. Moreover, we also incorporate the ID embeddings $\B{E}_{\C{S}}$ of items in the sequence $\C{S}$ with its text and image representations $\B{M}^t$ and $\B{M}^v$ using element-wise summation, and then feed the summed embeddings into the Transformer layers to obtain the sequence embeddings $\B{h}^t$ and $\B{h}^v$. Then, for the sequence $\C{S}$, the predicted probability distribution $\widehat{\B{y}}^{(\C{S})}$ of its potential target labels (\ie items) is defined as follows,
\begin{align}
    \widehat{\B{y}}^{(\C{S})} = \textrm{softmax}
    \big(\B{h}^{t}(\B{F}^t+\B{E})^{\top}
    +\B{h}^{v}(\B{F}^v+\B{E})^{\top}\big),
    \label{eq:ft1}
\end{align}
where $\widehat{\B{y}}^{(\C{S})}\in\mathbb{R}^{|\mathcal{I}|}$, $\B{E}\in\mathbb{R}^{|\mathcal{I}|\times d_0}$ denotes the ID embedding matrix of all items. $\B{F}^{t}, \B{F}^{v}\in\mathbb{R}^{|\mathcal{I}|\times d_0}$ denote the text and image modality embedding matrices of all items, which are obtained by the text encoder and image encoder in M$^2$SE. The loss function for model fine-tuning is defined as follows,
\begin{equation}
    \mathcal{L}_{\textrm{finetune}} = -\sum_{(\C{S}, i) \in \widetilde{\C{B}}} \log\big(\widehat{\B{y}}^{(\C{S})}(i)\big),
    \label{eq:s2}
\end{equation}
where $\widehat{\B{y}}^{(\C{S})}(i)$ denotes the predicted probability for the ground-truth label $i$ of the sequence $\C{S}$. By minimizing Eq.~\eqref{eq:s2}, we fine-tune the parameters of the pre-trained M$^2$SE network as well as the ID embedding of items.

%%%%%%%%%%%%%%%%%%%%%%%%%%%%%%%%%%%%%%%%%%%%%%%%%%%%%%%%%%%%%
\section{Experiments}

\subsection{Experimental Settings}

\subsubsection{Datasets}
\textcolor{black}{The experiments are conducted on the \textit{Poetry} category of GoodReads dataset~\cite{DBLP:conf/recsys/WanM18,DBLP:conf/acl/WanMNM19} and three categories of Amazon review dataset~\cite{ni2019justifying}, which includes \textit{Pantry}, \textit{Arts}, and \textit{Office}. These datasets provide the multimodal information of items. We use ``5-core'' subsets for experimental evaluation. }
Following~\cite{zhou2020s3,IJCAI-GCL4SR}, we convert each rating into an implicit feedback record.
On each dataset, we group interactions by users and construct the interaction sequence for each user by sorting her interactions in chronological order.
The statistics of the pre-processed experimental datasets are summarized in Table~\ref{tab:stats}.

\begin{table}
    \caption{Statistics of the experimental datasets. ``Avg. n'' denotes the average length of interaction sequences.}
  \centering
   % \small
  \begin{tabular}{l|cccc}
    \toprule
    Datasets & \#Users & \#Items & \#Inter. & Avg. n\\ 
    \midrule
    % Science &8822&17562&62083&7.04 \\
    \textcolor{black}{Poetry} & \textcolor{black}{11,314} & \textcolor{black}{11,231} & \textcolor{black}{108,114} &\textcolor{black}{9.55}\\
    Pantry & 13,614 & 7,670 &131,311 &9.65 \\
    Arts   & 32,216 & 52,557 &264,465 &8.21 \\
    Office & 68,224 & 59,705 &527,209 &7.73 \\
    \bottomrule
  \end{tabular}
  \label{tab:stats}
\end{table}

\subsubsection{Evaluation Settings}
Following~\cite{zhou2020s3,sun2019bert4rec}, we apply the \textit{leave-one-out} strategy to evaluate the performance of recommendation models in both pre-training and fine-tuning stages. Specifically, for each user, the last item of her interaction sequence is used for testing, the second last item is used for validation, and the remaining items are used for model training. The performance of a recommendation model is evaluated by two widely used metrics, \ie Recall@$K$ and Normalized Discounted Cumulative Gain@$K$ (respectively denoted by R@$K$ and N@$K$). $K$ is empirically set to 5, 10, and 20.
All evaluation metrics are computed on the whole candidate item set without negative sampling.

\textcolor{black}{For recommendation models, the approach to evaluating the
model’s predictive performance diverges from the direct application of training losses as
evaluation metrics. In our model, the primary function of training losses is to facilitate the
learning of the refined representation of users and items. For the evaluation phase, we adopt a
methodology that leverages cosine similarity calculations to gauge the alignment between user
embeddings and the embeddings of all potential candidate items. This process culminates in
the generation of a ranked list of items, prioritized according to their similarity to the user’s
profile. By employing this evaluative strategy, we ensure a comprehensive validation of our
next item prediction method, effectively bridging any potential gaps between the pre-training
tasks and the fine-tuning phase. This method not only aligns with established practices in
recommendation system evaluation but also offers a clear and direct measure of the model’s
utility in practical recommendation scenarios.}

\subsubsection{Baseline Methods}
We compare the proposed model with four groups of baseline methods:

1) \textit{General Recommendation Model:} 
\begin{itemize}
    \item \textbf{LightGCN}~\cite{he2020lightgcn} is one of the representative GNN-based recommendation methods.
\end{itemize}

2) \textit{Multimodal Recommendation Models:} 
\begin{itemize}
    \item \textbf{GRCN}~\cite{wei2020graph} is a graph-based multimodal recommendation model that refines the user-item interaction graph by pruning noisy edges;
    \item \textbf{DualGNN}~\cite{wang2021dualgnn} explicitly models the user's attention over different modalities.
\end{itemize}

3) \textit{Sequential Recommendation Models:} 
\begin{itemize}
    \item \textbf{SASRec}~\cite{kang2018self} is a directional self-attention method for next item prediction;
    \item \textbf{SINE}~\cite{tan2021sparse} uses a sparse-interest module that adaptively infers a sparse set of concepts and outputs multiple embeddings for each user;
    \item \textbf{CL4SRec}~\cite{xie2022contrastive} designs three data augmentation approaches to extract self-supervised signals to improve the sequential recommendation performance;
\end{itemize}

4) \textit{Sequential Recommendation Models with Side Information:}
\begin{itemize}
    \item \textbf{MV-RNN}~\cite{cui2018mv} is a multimodal sequential recommendation model that combines multimodal features at its input and applies a recurrent structure to dynamically capture users' interests; 
    \item \textcolor{black}{\textbf{FDSA}~\cite{zhang2019feature} models the transition patterns between items as well as features by separate self-attention blocks. We feed both text and image features into the model for fair comparison;}
    \item \textbf{$\textrm{S}^3$-Rec}~\cite{zhou2020s3} devises four self-supervised learning objectives based on the mutual information maximization principle;
    \item \textbf{DIF-SR}~\cite{xie2022decoupled} decouples attribute information and item representation from the calculation of attention;
    \item \textbf{UniSRec}~\cite{hou2022towards} leverages item text representations with an MoE-based adaptor and employs contrastive learning tasks to learn transferable sequence representations. For a fair comparison, we concatenate both raw texts and image-converted texts to train the model. We fine-tune the model under a transductive setting. This particular setting allows for the incorporation of both multimodal and ID representations associated with the items under consideration.
\end{itemize}

% \afterpage{
%     \clearpage % End the current page
%     \thispagestyle{empty} % Apply the empty page style to the next page
\begin{table*}
    \caption{The overall performance achieved by different methods. The best results are in \textbf{boldface}, and the second best results are \underline{underlined}. * denotes MP4SR surpasses the best baseline using a paired t-test ($p < 0.01$).}
    \centering
    % \small
    % \resizebox{0.8\textwidth}{!}{
    \begin{tabular}{l|l| c c c| c c c }
    \toprule
    Dataset & Model & R@5 & R@10 & R@20 & N@5 & N@10 & N@20 \\
    \midrule
    \multirow{12}{*}{\textcolor{black}{Poetry}}
    &\textcolor{black}{LightGCN}&\textcolor{black}{0.1126}&\textcolor{black}{0.1763}&\textcolor{black}{0.2592}&\textcolor{black}{0.0737}&\textcolor{black}{0.0941}&\textcolor{black}{0.1150}\\
    &\textcolor{black}{GRCN}&\textcolor{black}{0.1031}&\textcolor{black}{0.1712}&\textcolor{black}{0.2571}&\textcolor{black}{0.0649}&\textcolor{black}{0.0868}&\textcolor{black}{0.1084}\\
    &\textcolor{black}{DualGNN}&\textcolor{black}{0.1033}&\textcolor{black}{0.1691}&\textcolor{black}{0.2553}&\textcolor{black}{0.0641}&\textcolor{black}{0.0852}&\textcolor{black}{0.1070}\\
    &\textcolor{black}{SASRec}&\textcolor{black}{0.1120}&\textcolor{black}{0.1886}&\textcolor{black}{0.2793}&\textcolor{black}{0.0654}&\textcolor{black}{0.0900}&\textcolor{black}{0.1129}\\
    &\textcolor{black}{SINE}&\textcolor{black}{0.0972}&\textcolor{black}{0.1703}&\textcolor{black}{0.2728}&\textcolor{black}{0.0593}&\textcolor{black}{0.0828}&\textcolor{black}{0.1086}\\
    &\textcolor{black}{CL4SRec}&\textcolor{black}{0.1044}&\textcolor{black}{0.1936}&\textcolor{black}{0.2902}&\textcolor{black}{0.0602}&\textcolor{black}{0.0889}&\textcolor{black}{0.1133}\\
    &\textcolor{black}{MV-RNN}&\textcolor{black}{0.0803}&\textcolor{black}{0.1355}&\textcolor{black}{0.2192}&\textcolor{black}{0.0495}&\textcolor{black}{0.0667}&\textcolor{black}{0.0878}\\
    &\textcolor{black}{FDSA}&\textcolor{black}{0.1224}&\textcolor{black}{0.1959}&\textcolor{black}{0.2873}&\textcolor{black}{0.0788}&\textcolor{black}{0.1024}&\textcolor{black}{\underline{0.1254}}\\
    &\textcolor{black}{$\textrm{S}^3$-Rec}&\textcolor{black}{\underline{0.1227}}&\textcolor{black}{0.1877}&\textcolor{black}{0.2723}&\textcolor{black}{\textbf{0.0816}}&\textcolor{black}{\underline{0.1025}}&\textcolor{black}{0.1238}\\
    &\textcolor{black}{DIF-SR}&\textcolor{black}{0.1225}&\textcolor{black}{\underline{0.1992}}&\textcolor{black}{\underline{0.2982}}&\textcolor{black}{0.0713}&\textcolor{black}{0.0959}&\textcolor{black}{0.1209}\\
    &\textcolor{black}{UniSRec}&\textcolor{black}{0.1120}&\textcolor{black}{0.1890}&\textcolor{black}{0.2783}&\textcolor{black}{0.0636}&\textcolor{black}{0.0883}&\textcolor{black}{0.1109}\\
    \cline{2-8}
    &\textcolor{black}{\textbf{MP4SR}}&\textcolor{black}{\textbf{0.1322*}}&\textcolor{black}{\textbf{0.2109*}}&\textcolor{black}{\textbf{0.3057}}&\textcolor{black}{\underline{0.0807}}&\textcolor{black}{\textbf{0.1059*}}&\textcolor{black}{\textbf{0.1299*}}\\
    \midrule
    \multirow{12}{*}{Pantry}
    &LightGCN&0.0270&0.0460&0.0774&0.0176&0.0236&0.0315\\
    &GRCN&0.0365&0.0552&0.0856&\underline{0.0229}&\underline{0.0289}&0.0366\\
    &DualGNN&0.0321&0.0485&0.0739&0.0202&0.0254&0.0318\\
    &SASRec&0.0277&0.0457&0.0722&0.0147&0.0204&0.0271\\
    &SINE&0.0297&0.0534&0.0873&0.0167&0.0243&0.0329\\
    &CL4SRec&0.0289&0.0487&0.0796&0.0173&0.0236&0.0314\\
    &MV-RNN&0.0157&0.0276&0.0467&0.0101&0.0134&0.0184\\
    &FDSA&0.0235&0.0357&0.0588&0.0155&0.0194&0.0252\\
    &$\textrm{S}^3$-Rec&0.0315&0.0535&0.0845&0.0187&0.0257&0.0335\\
    &DIF-SR&0.0300&0.0473&0.0736&0.0163&0.0219&0.0284\\
    & UniSRec &\underline{0.0381}&\underline{0.0616}&\underline{0.0965}&0.0212&0.0287&\underline{0.0374} \\
    \cline{2-8}
    &\textbf{MP4SR}*&\textbf{0.0405}*&\textbf{0.0673}*&\textbf{0.1040}*&\textbf{0.0235}*&\textbf{0.0321}*&\textbf{0.0414}*\\
    
    \midrule
    \multirow{12}{*}{Arts}
    &LightGCN	&0.0543&0.0726&0.0967&0.0381&0.044&0.0501 \\
    &GRCN	&0.0546&0.0741&0.0999&0.0386&0.0448&0.0513 \\
    &DualGNN	&0.0596&0.0788&0.1033&0.0433&0.0495&0.0557 \\
    &SASRec	&0.0704&0.0910&0.1125&0.0442&0.0509&0.0563 \\
    &SINE	&0.0667&0.0935&0.1237&0.0404&0.0491&0.0567 \\
    &CL4SRec	&0.0670&0.0899&0.1162&0.0410&0.0484&0.0550 \\
    &MV-RNN	&0.0299&0.0446&0.0661&0.0184&0.0232&0.0283 \\
    &FDSA	&0.0635&0.0772&0.0948&\underline{0.0501}&0.0545&0.0589 \\
    &$\textrm{S}^3$-Rec	&0.0715&0.0961&0.1250&0.0467&\underline{0.0546}&\underline{0.0619} \\
    &DIF-SR	&0.0712&0.0899&0.1126&0.0449&0.0510&0.0567 \\
    &UniSRec &\underline{0.0747}&\underline{0.1026}&\underline{0.1339}&0.0447&0.0537&0.0616 \\
    \cline{2-8}
    &\textbf{MP4SR}	&\textbf{0.0854}*&\textbf{0.1184}*&\textbf{0.1570}*&\textbf{0.0531}*&\textbf{0.0637}*&\textbf{0.0735}* \\
    \midrule
    \multirow{12}{*}{Office}
    &LightGCN&0.0325&0.0518&0.0752&0.0219&0.0281&0.0339\\
    &GRCN&0.0556&0.0714&0.0911&0.0408&0.0460&0.0509\\
    &DualGNN&0.0518&0.0661&0.0843&0.0385&0.0431&0.0477\\
    &SASRec&0.0841&0.1025&0.1222&0.0558&0.0617&0.0667\\
    &SINE&0.0837&0.1059&0.1305&0.0546&0.0618&0.0680\\
    &CL4SRec&0.0799&0.1016&0.1256&0.0531&0.0602&0.0662\\
    &MV-RNN&0.0259&0.0416&0.0641&0.0159&0.0210&0.0266\\
    &FDSA&0.0703&0.0832&0.0997&0.0574&0.0616&0.0657\\
    &$\textrm{S}^3$-Rec&0.0823&0.1027&0.1254&\underline{0.0575}&\underline{0.0641}&\underline{0.0698}\\
    &DIF-SR&0.0857&0.1039&0.1241&0.0561&0.0620&0.0671\\
    & UniSRec &\underline{0.0868}&\underline{0.1085}&\underline{0.1322}&0.0543&0.0613&0.0673\\
    \cline{2-8} &\textbf{MP4SR}&\textbf{0.0968}*&\textbf{0.1206}*&\textbf{0.1480}*&\textbf{0.0721}*&\textbf{0.0797}*&\textbf{0.0866}*\\
    \bottomrule
    \end{tabular}
    \label{tab:overall}
    
\end{table*}
%     \clearpage % End the table page
% }

\subsubsection{Implementation Details}
The proposed method is implemented by Pytorch~\cite{paszke2019pytorch} and an open-source recommendation framework, RecBole~\cite{zhao2021recbole}. The Adam optimizer~\cite{kingma2014adam} is used to learn model parameters. Following~\cite{zhou2020s3}, we set the maximum sequence length to $50$. Our training phase consists of two stages: pre-training and fine-tuning. The learned parameters from the pre-training stage are used to initialize the M$^2$SE network in the fine-tuning stage. Both the pre-training and fine-tuning are performed on the same dataset to obtain the final recommendation results.

In the multimodal feature extraction stage of MP4SR, 
the pre-trained Sentence-BERT model maps every sentence of text descriptions or a group of word tokens extracted from an image into a $768$-dimensional dense vector, \ie $d=768$. For each item, we consider up to $10$ sentences and $10$ images. 

For pre-training MP4SR, we set the learning rate to 0.001, the batch size to 1024, and the number of experts $O$ to 8 on all datasets. In addition, we set $\rho$, $\tau$, and $\lambda$ to 0.2, 0.07, and 0.01, respectively. The attention dimension $d_a$ and embedding dimension $d_0$ are fixed to $64$. The proposed model is pre-trained for $300$ epochs. 

For fine-tuning MP4SR and training baseline methods, 
we apply grid search to identify the best hyper-parameter settings based on the validation data for each method. The search space is as follows: learning rate in $\{0.0001, 0.0005, 0.001\}$, batch size in $\{256, 512, 1024\}$, and weight decay in $\{0.0001, 0.0005, 0.001\}$. 
For a fair comparison, the hyper-parameters of Transformer layers are kept identical for MP4SR and transformer-based baselines. Specifically, the number of attention heads and the number of self-attention blocks are set to 2. The remaining hyper-parameters for baseline methods follow the original papers. Additionally, we adopt an early stopping strategy, \ie we apply premature stopping if R@20 on the validation data does not increase for 10 epochs.

%%%%%%%%%%%%%%%%%%%%%%%%%%%%%%%%%%%%%%
\subsection{Performance Comparison}

We summarize the overall performance comparison results in Table~\ref{tab:overall}, from which we have the following observations:
\begin{itemize}
    \item The multimodal recommendation methods (\ie GRCN and DualGNN) consistently outperform the general recommendation model (\ie LightGCN) for most datasets. This suggests that leveraging the multimodal information of items can effectively enhance recommendation performance.
    \item Sequential recommendation models generally perform better than non-sequential recommendation models, by capturing users' sequential behavior patterns. 
    Notably, an exception is observed in the Pantry dataset, where the non-sequential model GRCN, which utilizes multimodal data, achieves better performance than sequential baseline methods.
    This may be attributed to the fact that the multimodal features of items in this dataset are more informative than those in the other two datasets. 
    In addition, it is evident that CL4SRec outperforms other sequential baselines, demonstrating the effectiveness of leveraging data augmentation to enhance the performance of sequential recommendation.
    \item $\textrm{S}^3$-Rec or UniSRec usually surpasses other sequential recommendation baselines, highlighting the effectiveness of using self-supervised signals and side information for pre-training sequential recommendation models.
    \item The proposed MP4SR model consistently outperforms all baseline methods by a significant margin. 
    This is attributed to the utilization of two pre-training objectives to capture the correlation of multimodal data with user behaviors, thereby improving the generalization capabilities of sequential recommendation models and resulting in the best overall performance.
\end{itemize}

%%%%%%%%%%%%%%%%%%%%%%%%%%%%%%%%%%%%%%
\subsection{Cold-start Performance}
\begin{table}
    \caption{The performance of cold-items achieved by CLCRec, MASR, MP4SR\textsubscript{w/o Pre-train}, and MP4SR. }
    % \small
    \centering
    \begin{tabular}{l|l|c c| c c}
    \toprule
    Dataset & Model & R@10& R@20 & N@10 & N@20\\
    \midrule
    \multirow{4}{*}{Pantry}
    &MASR&0.0111&0.0119&0.0064&0.0066 \\
    &CLCRec &0.0166&0.0251&0.0081&0.0103\\
    &MP4SR\textsubscript{w/o Pre-train} &0.0257&0.0337&0.0119&0.0139\\
    &MP4SR&\textbf{0.0360}&\textbf{0.0491}&\textbf{0.0176}&\textbf{0.0208} \\
    \midrule
    \multirow{4}{*}{Arts}
    &MASR&0.0136&0.0165&0.0080&0.0090 \\
    &CLCRec &0.0178&0.0239&0.0101&0.0116\\
    &MP4SR\textsubscript{w/o Pre-train} &0.0314&0.0455&0.0145&0.0181\\
    &MP4SR&\textbf{0.0403}&\textbf{0.0552}&\textbf{0.0191}&\textbf{0.0229}\\
    \midrule
    \multirow{4}{*}{Office}
    &MASR&0.0079&0.0094&0.0050&0.0054 \\
    &CLCRec&0.0094&0.0120&0.0049&0.0056 \\
    &MP4SR\textsubscript{w/o Pre-train} &0.0128&0.0183&0.0061&0.0074\\
    &MP4SR &\textbf{0.0235}&\textbf{0.0312}&\textbf{0.0117}&\textbf{0.0136}\\
    \bottomrule
    \end{tabular}
    \label{tab:cold}
\end{table}
To validate the effectiveness of our model for the cold-start recommendation, we include the following methods for evaluation alongside \textbf{MP4SR}\textsubscript{w/o Pre-train} and \textbf{MP4SR}: 
\begin{itemize}
    \item \textbf{CLCRec}~\cite{wei2021contrastive}: this method explores the mutual dependency between item multimodal features and collaborative representations to alleviate the cold-start item problem.
    \item \textbf{MASR}~\cite{hu2022memory}: authors construct two memory banks to store historical user sequences and a retriever-copy network to search for similar sequences to enhance the recommendation performance for cold-start items.
\end{itemize}

In our experiments, counting all items in the training set, we categorize those that appear less than 10 times as cold items, and the rest as warm items. CLCRec, MASR, and MP4SR\textsubscript{w/o Pre-train} are trained based on the full dataset, including both cold and warm items, and are evaluated based on user sequences that take cold items as the target item for prediction. MP4SR is first pre-trained on warm items, followed by fine-tuning using the entire dataset. Its performance is evaluated in the same manner as the other three baselines. Given that cold items lack sufficient interaction data, item ID embeddings are excluded during fine-tuning.

Table~\ref{tab:cold} shows the performance achieved by CLCRec, MASR, MP4SR\textsubscript{w/o Pre-train}, and MP4SR on cold items.
We can note that both MP4SR\textsubscript{w/o Pre-train} and MP4SR outperform two baseline methods, illustrating the effectiveness of using multimodal information to alleviate the cold-start item problem. Overall, MP4SR performs the best by a substantial margin across all evaluation metrics. This result suggests that cold items can benefit more from self-supervised multimodal pre-training tasks that leverage items with more interactions.

%%%%%%%%%%%%%%%%%%%%%%%%%%%%%%%%%%%%%%
\subsection{Ablation Study}
\begin{table}
    \caption{The performance of MP4SR and its variants on Pantry and Office datasets. }
    % \small
    \centering
    \begin{tabular}{l|l|c c| c c}
    \toprule
    Dataset & Model & R@10& R@20 & N@10 & N@20\\
    \midrule
    \multirow{9}{*}{Pantry}
     &MP4SR\textsubscript{ResNet} &0.0647&0.1007&0.0317&0.0408\\
     &MP4SR\textsubscript{w/o NIP} &0.0501&0.0816&0.0245	&0.0324\\
     &MP4SR\textsubscript{w/o CMCL} &0.0662& 0.1030&0.0310& 0.0403\\
     &MP4SR\textsubscript{w/o C-Mixup} &0.0649&0.1014 &0.0307&0.0399\\
     &MP4SR\textsubscript{w/o Pre-train} &0.0595&0.0920&0.0286&0.0369\\
     &MP4SR\textsubscript{w/o Proj}&0.0630&0.1001&0.0300&0.0393\\
     &MP4SR\textsubscript{E2E} &0.0605&0.0937&0.0290&0.0373 \\
     &\textcolor{black}{MP4SR\textsubscript{Shared}} &\textcolor{black}{0.0644}&\textcolor{black}{0.0987}&\textcolor{black}{0.0309}&\textcolor{black}{0.0395} \\
     \cline{2-6}
     &MP4SR &\textbf{0.0673} &\textbf{0.1040}&\textbf{0.0321}&\textbf{0.0414}\\
    \midrule
    \multirow{9}{*}{Office}
     &MP4SR\textsubscript{ResNet} &0.1159&0.1435&0.0749&0.0818\\
     &MP4SR\textsubscript{w/o NIP} &0.1062&0.1302&0.0637&0.0697\\
     &MP4SR\textsubscript{w/o CMCL} &0.1095&0.1349&0.0649&0.0713\\
     &MP4SR\textsubscript{w/o C-Mixup} &0.1192&\textbf{0.1480}&0.0768&0.0841\\
     &MP4SR\textsubscript{w/o Pre-train} &0.1094	&0.1335&0.0665&0.0726\\
     &MP4SR\textsubscript{w/o Proj}&0.1177&0.1456&0.0724&0.0794\\
     &MP4SR\textsubscript{E2E}&0.1013&0.1243&0.0589&0.0647 \\
    &\textcolor{black}{MP4SR\textsubscript{Shared}} &\textcolor{black}{0.0899}&\textcolor{black}{0.1112}&\textcolor{black}{0.0717}&\textcolor{black}{0.0779} \\
    \cline{2-6}
    &MP4SR &\textbf{0.1206}&\textbf{0.1480}&\textbf{0.0797}&\textbf{0.0866}\\
    \bottomrule
    \end{tabular}
    \label{tab:ablation}
\end{table}
To study the contribution of each component of MP4SR, we consider the following variants of MP4SR for evaluation: 
\begin{itemize}
    \item \textbf{MP4SR}\textsubscript{ResNet}: we use ResNet to extract features of item images, instead of converting an item image into keywords; 
    % 2) \textbf{MP4SR}\textsubscript{w/o Seq Dropout}: we remove the sequence random dropout module in M$^2$SE; 
    \item \textbf{MP4SR}\textsubscript{w/o NIP}: we remove the modality-wise next item prediction losses in the pre-training stage; 
    \item \textbf{MP4SR}\textsubscript{w/o CMCL}: we remove the cross-modality contrastive losses in the pre-training stage;
    \item \textbf{MP4SR}\textsubscript{w/o C-Mixup}: we remove the complementary sequence mixup module in M$^2$SE;
    % is removed and sequences only contain unimodal information;
    \item \textbf{MP4SR}\textsubscript{w/o Pre-train}: we remove the pre-training tasks and train the proposed model from scratch based on the multimodal fine-tuning setting;
    \item \textbf{MP4SR}\textsubscript{w/o Proj}: we remove the two projection heads and calculate the pre-training contrastive losses on $\B{h}^t$ and $\B{h}^v$;
    \item \textbf{MP4SR}\textsubscript{E2E}: we optimize the proposed model in an end-to-end manner by summing up the pre-training loss~$\mathcal{L}_{\textrm{pre-train}}$ and the finetuning loss~$\mathcal{L}_{\textrm{finetune}}$.
    \item \textbf{MP4SR\textsubscript{Shared}}: \textcolor{black}{we share the model parameters for both the text and image modality encoders.}
\end{itemize}

Table~\ref{tab:ablation} presents the performance of MP4SR and its variants on Pantry and Office datasets. It shows that each proposed component of MP4SR consistently improves recommendation performance. 
The modality-wise next item prediction losses are particularly important for pre-training MP4SR for sequential recommendation. Omitting them results in a significant decline in performance.
This is likely due to the fact that 
% the model optimizes next item prediction as its main objective during both pre-training and fine-tuning stages.
the model's primary objective is the optimization of next item prediction throughout both pre-training and fine-tuning stages.
Furthermore, cross-modality contrastive losses are more effective in improving recommendation performance on the Office dataset compared to the Pantry dataset, indicating that items in the Office dataset provide more training signals to align various modalities.
Additionally, we note that the recommendation performance decreases significantly when pre-training tasks are eliminated, which further validates the effectiveness of applying pre-training for the multimodal sequential recommendation.
Also, projection heads facilitate the calculation of contrastive losses by mapping each sequence representation into a common semantic space. If they are removed, the performance is negatively affected.
Next, if the model is trained end-to-end by combining the $\mathcal{L}_{\textrm{pre-train}}$ and $\mathcal{L}_{\textrm{finetune}}$, the performance deteriorates. This is because pre-training losses aim to learn interactions across different modalities, whereas finetuning losses prioritize recommendation tasks using cross-entropy losses. If these are optimized together, the model struggles to converge to the optimal solution for the recommendation task.
\textcolor{black}{Finally, we evaluate the model's performance when the model parameters are shared between the text and image modality encoders. The performance deteriorates in this scenario. Despite converting images to text in this study, it is evident that images and texts still carry distinct information. Hence, employing separate modality encoders to encode features into different spaces will be advantageous for our pre-training and fine-tuning framework.}

\subsection{Parameter Sensitivity Study}

In this experiment, we study the impact of four hyper-parameters, including the $\lambda$ to balance between modality-wise next item prediction loss and cross-modality contrastive loss, the number of tokens retrieved for each image $N$, the number of experts used in the MoE architecture $O$, and the random dropout probability of a sequence $\rho$. We conduct experiments on Pantry and Office and report R@20 for comparison.

\subsubsection{Impact of $\lambda$}
\begin{figure}
\centering
\subfloat[$\lambda$-R@20]{\includegraphics[width=0.35\textwidth]{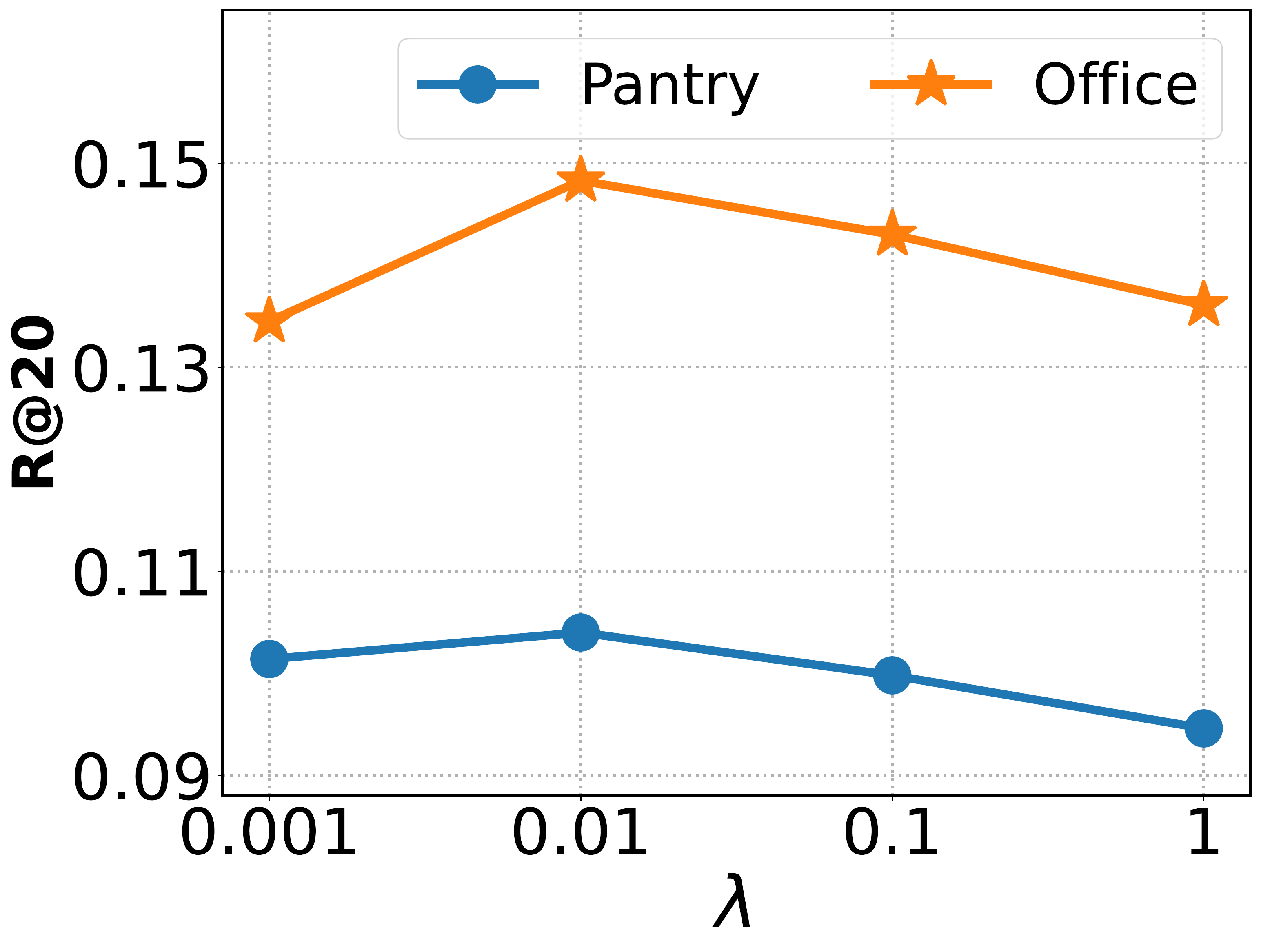}}
\qquad
\subfloat[$\lambda$-N@20]{\includegraphics[width=0.35\textwidth]{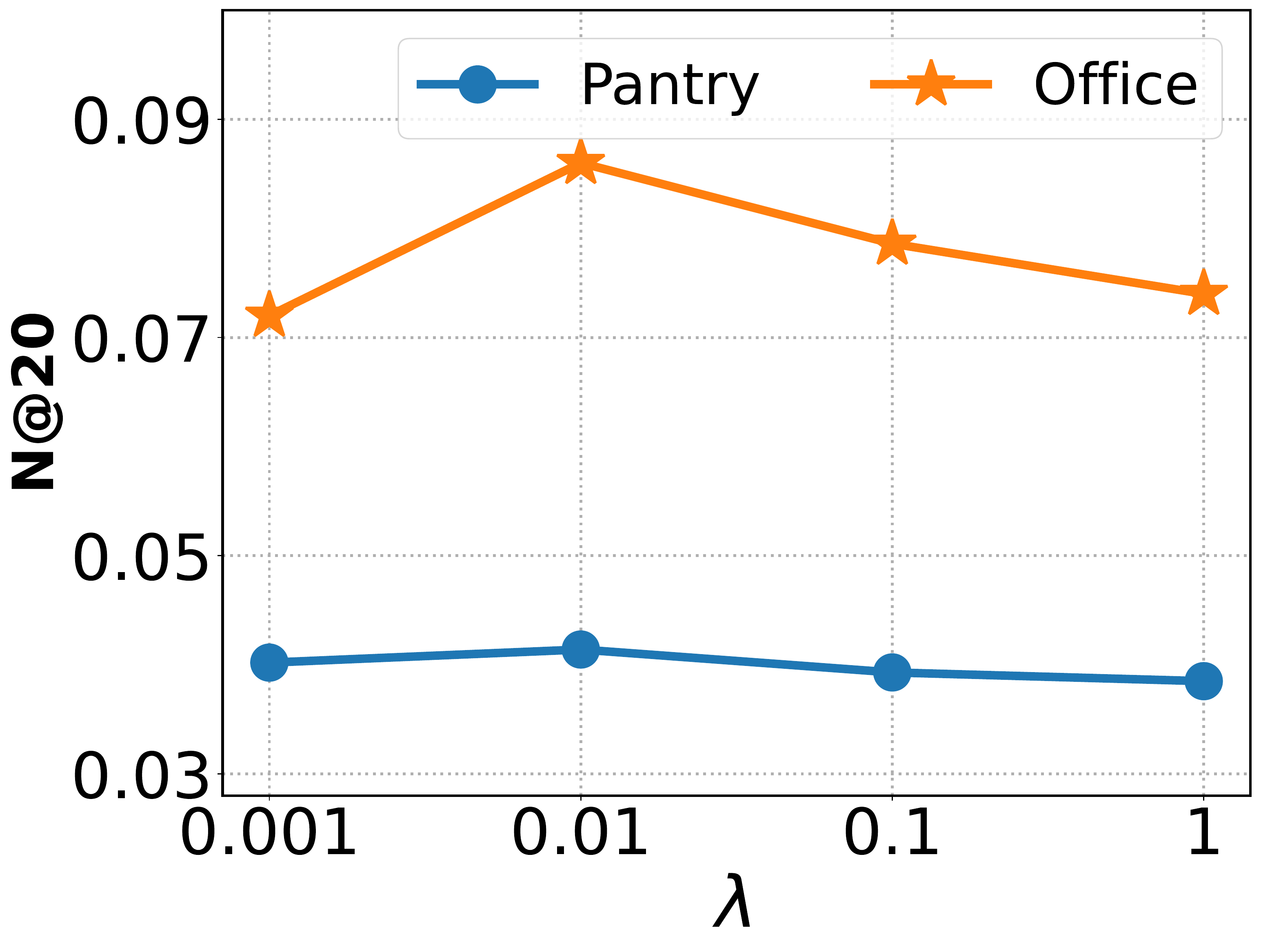}}
\caption{The performance trends of MP4SR with respect to different settings of $\lambda$ on Pantry and Office datasets.}
\label{fig:parameters-lam}
\end{figure}

The performance comparison using different values of $\lambda$ is shown in Figure~\ref{fig:parameters-lam}. We vary $\lambda$ in $\{0.001, 0.01, 0.1, 1.0\}$. We can note that the best performance is achieved when $\lambda$ is set to $0.01$. Recommendation performance is compromised with either a large or a small $\lambda$.

\subsubsection{Impact of $N$}
\begin{figure}
\centering
\subfloat[$N$-R@20]{\includegraphics[width=0.35\textwidth]{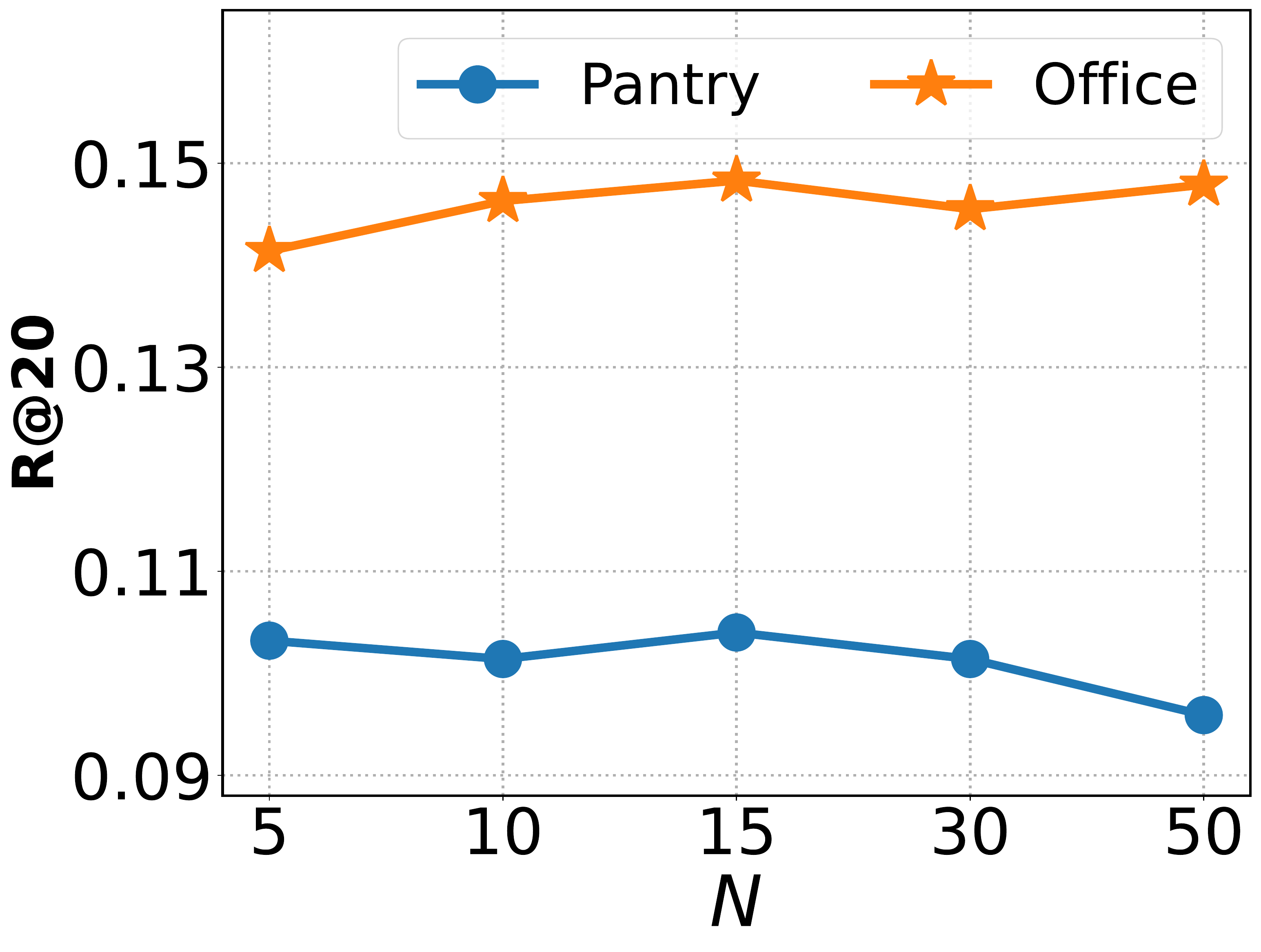}}
\qquad
\subfloat[$N$-N@20]{\includegraphics[width=0.35\textwidth]{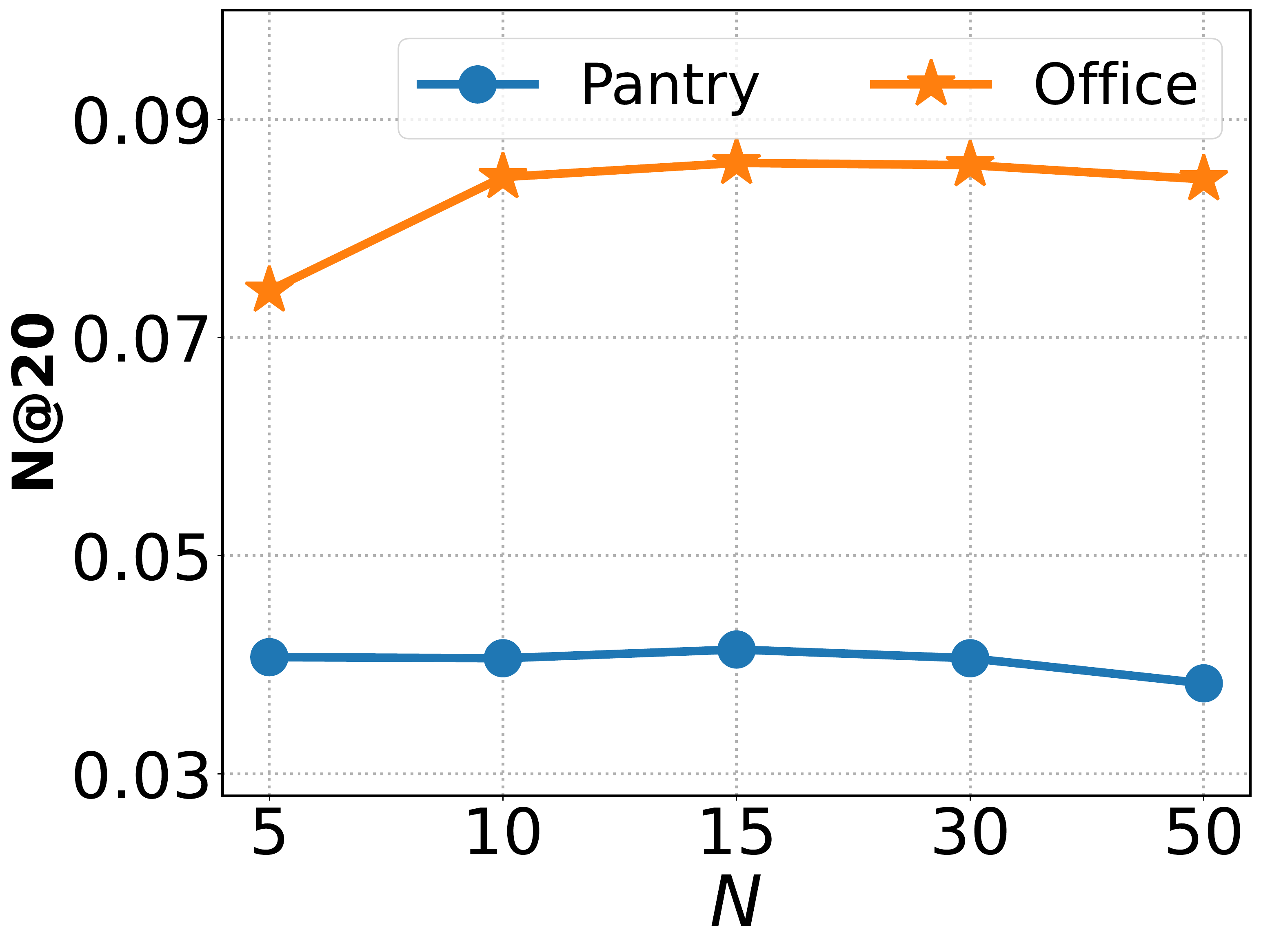}}
\caption{The performance trends of MP4SR with respect to different settings of $N$ on Pantry and Office datasets.}
\label{fig:parameters-n}
\end{figure}

The number of tokens retrieved for each image $N$ is varied in $\{5, 10, 15, 30, 50\}$. As shown in Figure~\ref{fig:parameters-n}, $15$ word tokens per image appear to be the optimal setting for both datasets. Insufficient information from images is captured when fewer tokens are used, while excessive token usage usually introduces noise.

\subsubsection{Impact of $O$}
\begin{figure*}
\centering
\subfloat[$O$-R@20]{\includegraphics[width=0.35\textwidth]{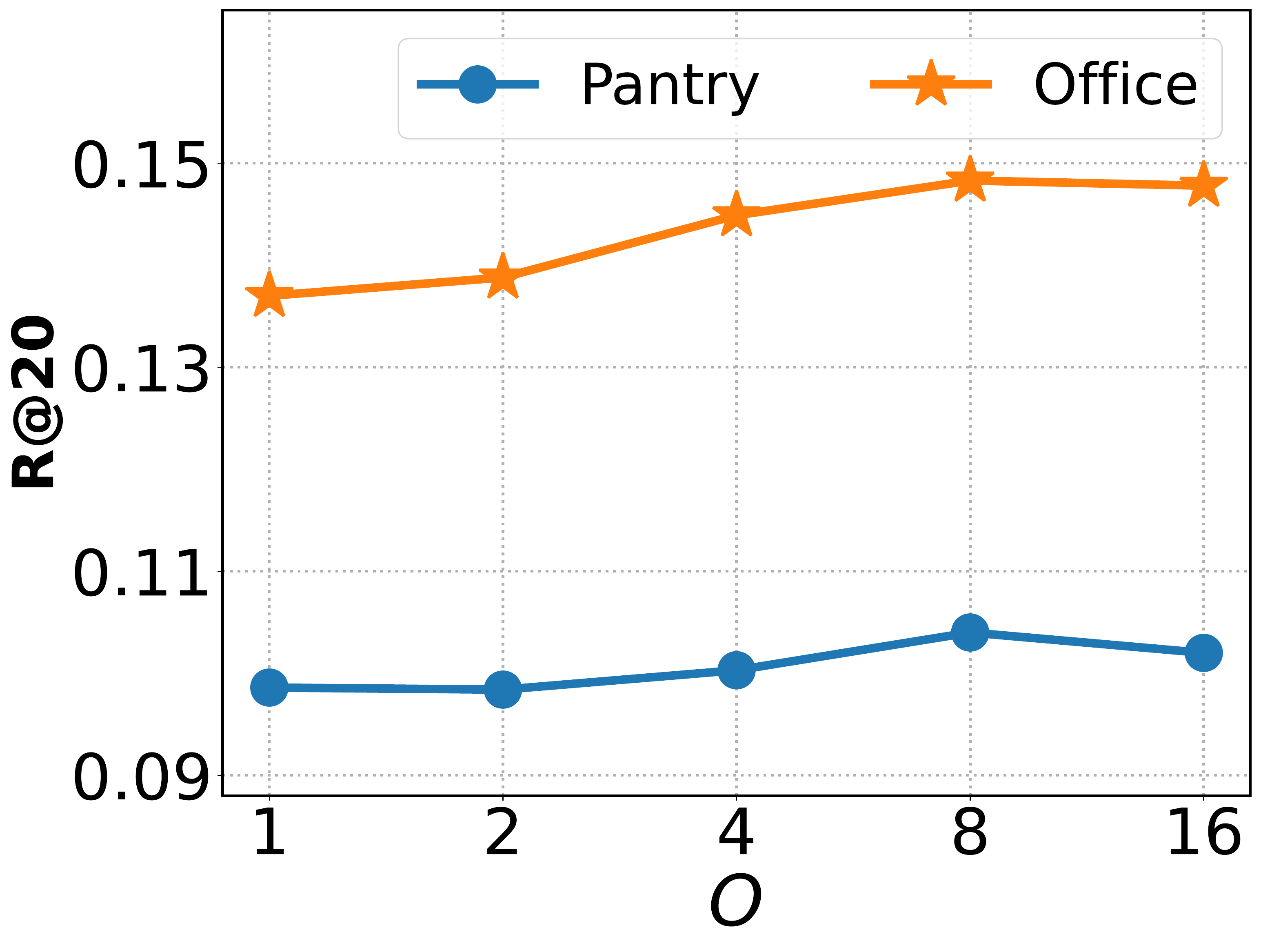}}
\qquad
\subfloat[$O$-N@20]{\includegraphics[width=0.35\textwidth]{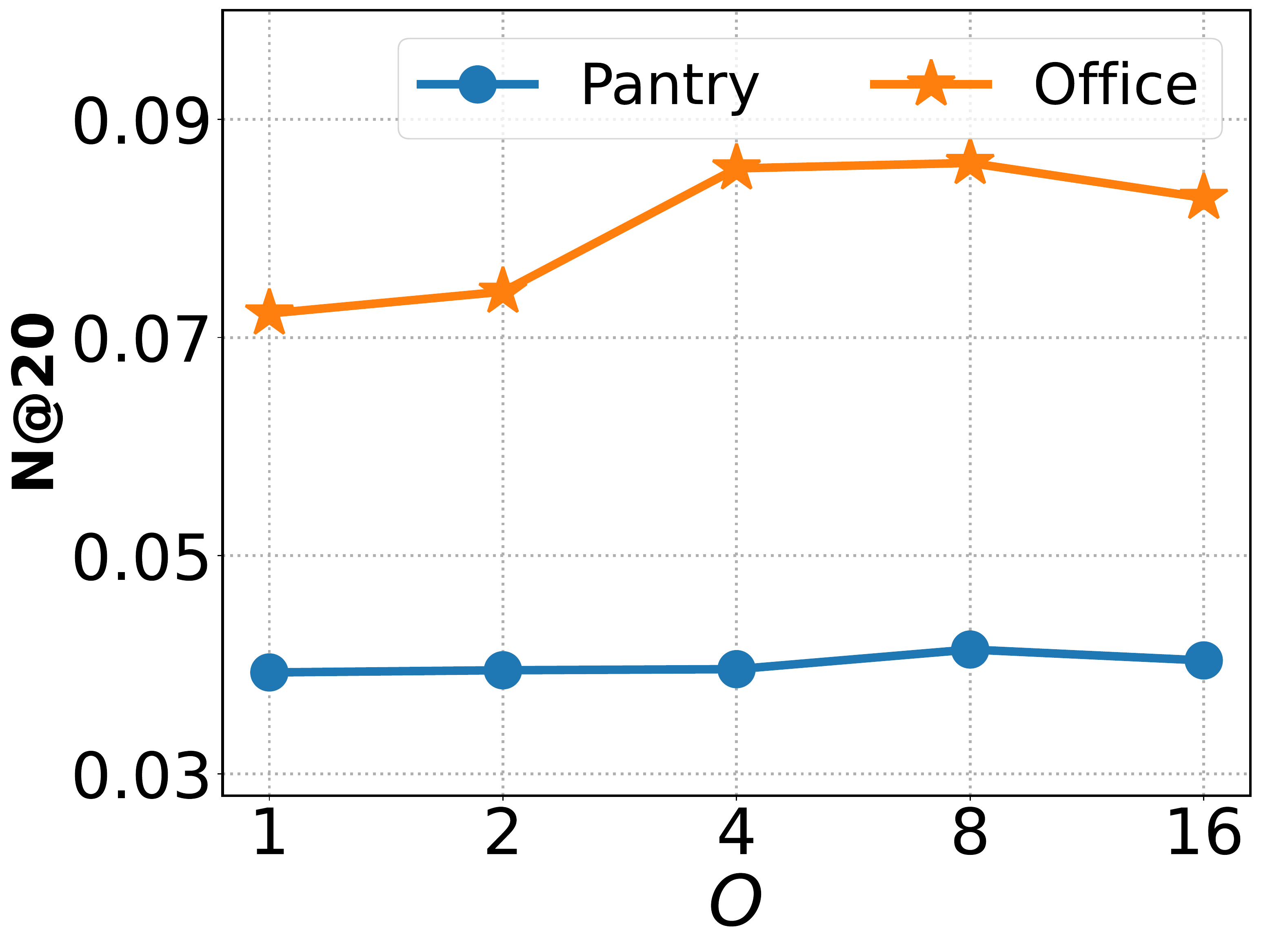}}
\caption{The performance trends of MP4SR with respect to different settings of $O$ on Pantry and Office datasets.}
\label{fig:parameters-o}
\end{figure*}

The number of experts used in the MoE architecture $O$ is chosen from $\{1, 2, 4, 8, 16\}$. From Figure~\ref{fig:parameters-o}, we can notice the results with respect to $O$ are consistent on both datasets. The best performance is achieved when $O$ is set to 8. However, the further increase of $O$ does not help with the performance.

\subsubsection{Impact of $\rho$}
\begin{figure*}
\centering
\subfloat[$\rho$-R@20]{\includegraphics[width=0.35\textwidth]{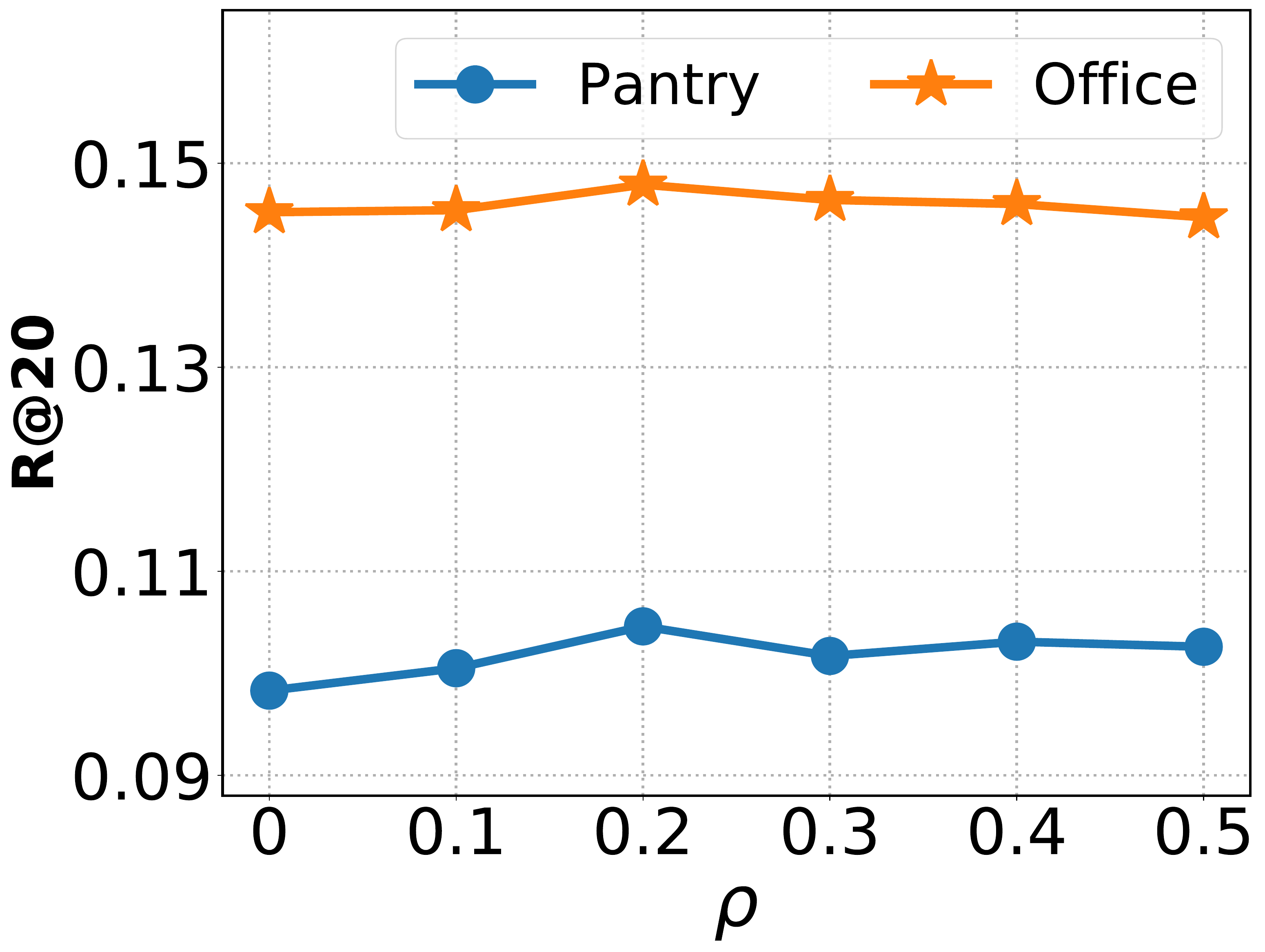}}
\qquad
\subfloat[$\rho$-N@20]{\includegraphics[width=0.35\textwidth]{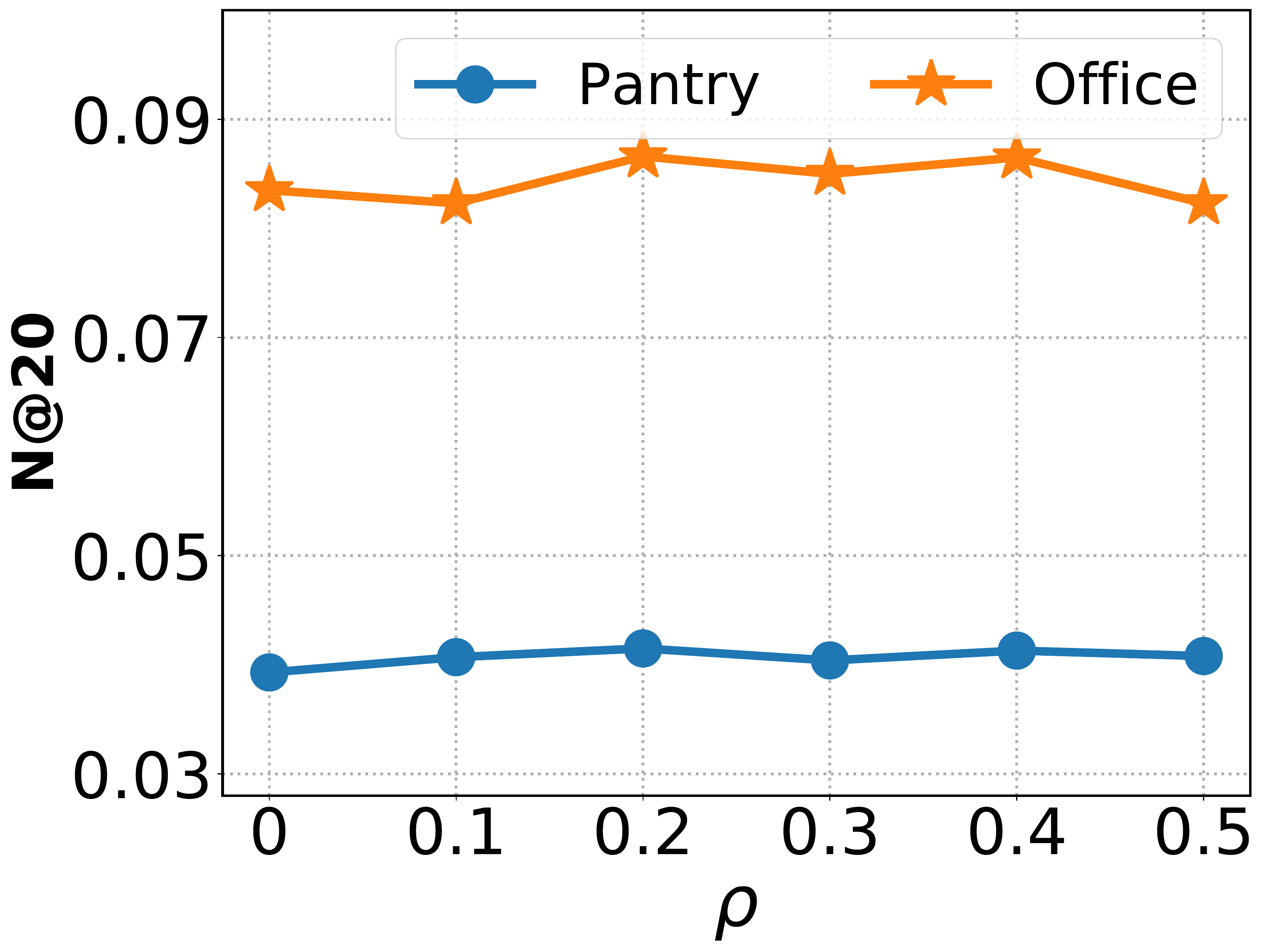}}
\
\caption{The performance trends of MP4SR with respect to different settings of $\rho$ on Pantry and Office datasets.}
\label{fig:parameters-rho}
\end{figure*}

As shown in Figure~\ref{fig:parameters-rho}, we examine the model performance with different dropout probabilities of the user sequence $\rho$, which ranges from $0$ to $0.5$ with a step size of $0.1$. We can observe that the model performance is relatively stable with the change in the random dropout probability.

%%%%%%%%%%%%%%%%%%%%%%%%%%%%%%%%%%%%
\subsection{Performance on Different User Groups}
\begin{figure}
\centering
\subfloat[Pantry-Recall]{\includegraphics[width=0.35\textwidth]{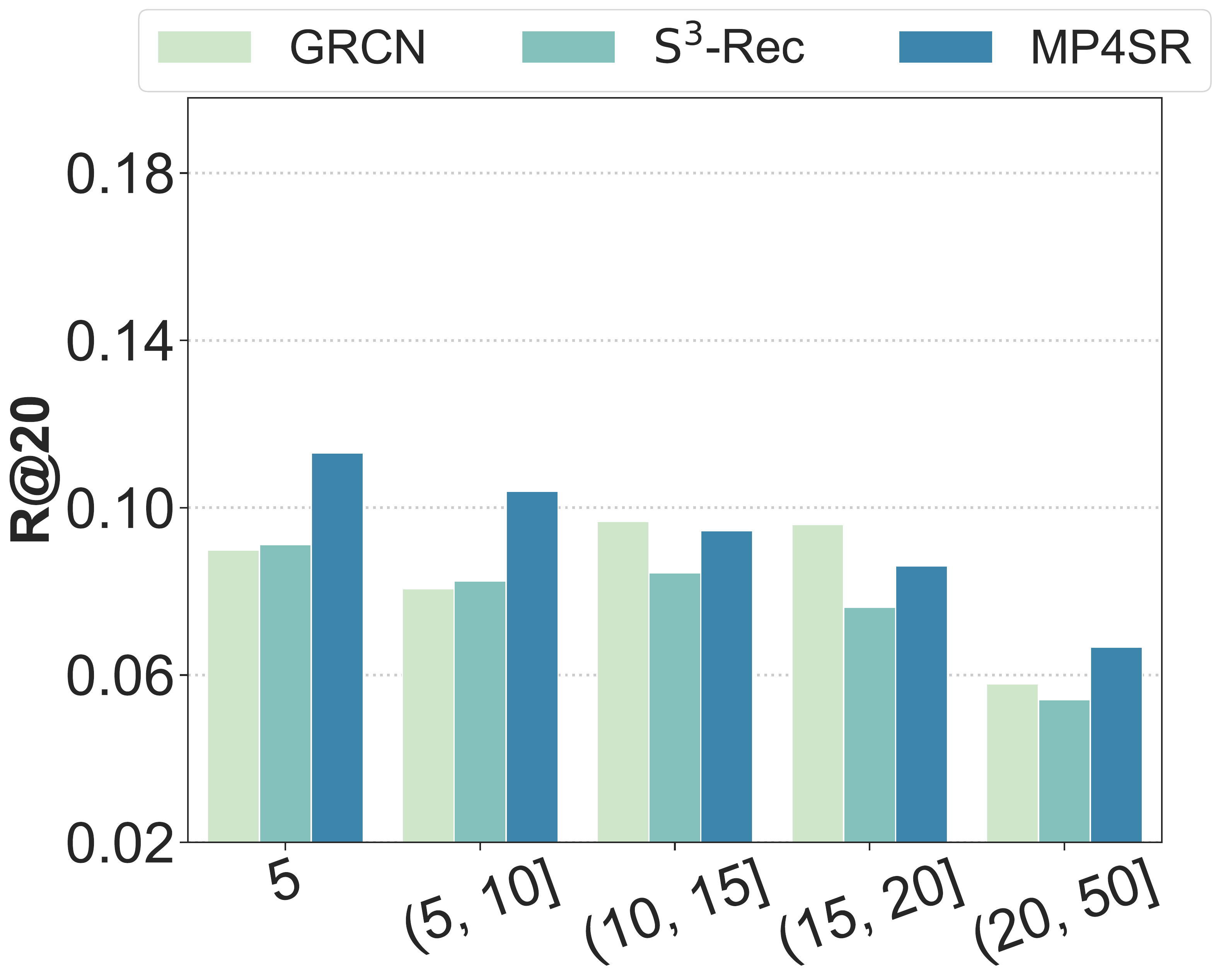}}
\qquad
\subfloat[Pantry-NDCG]{\includegraphics[width=0.35\textwidth]{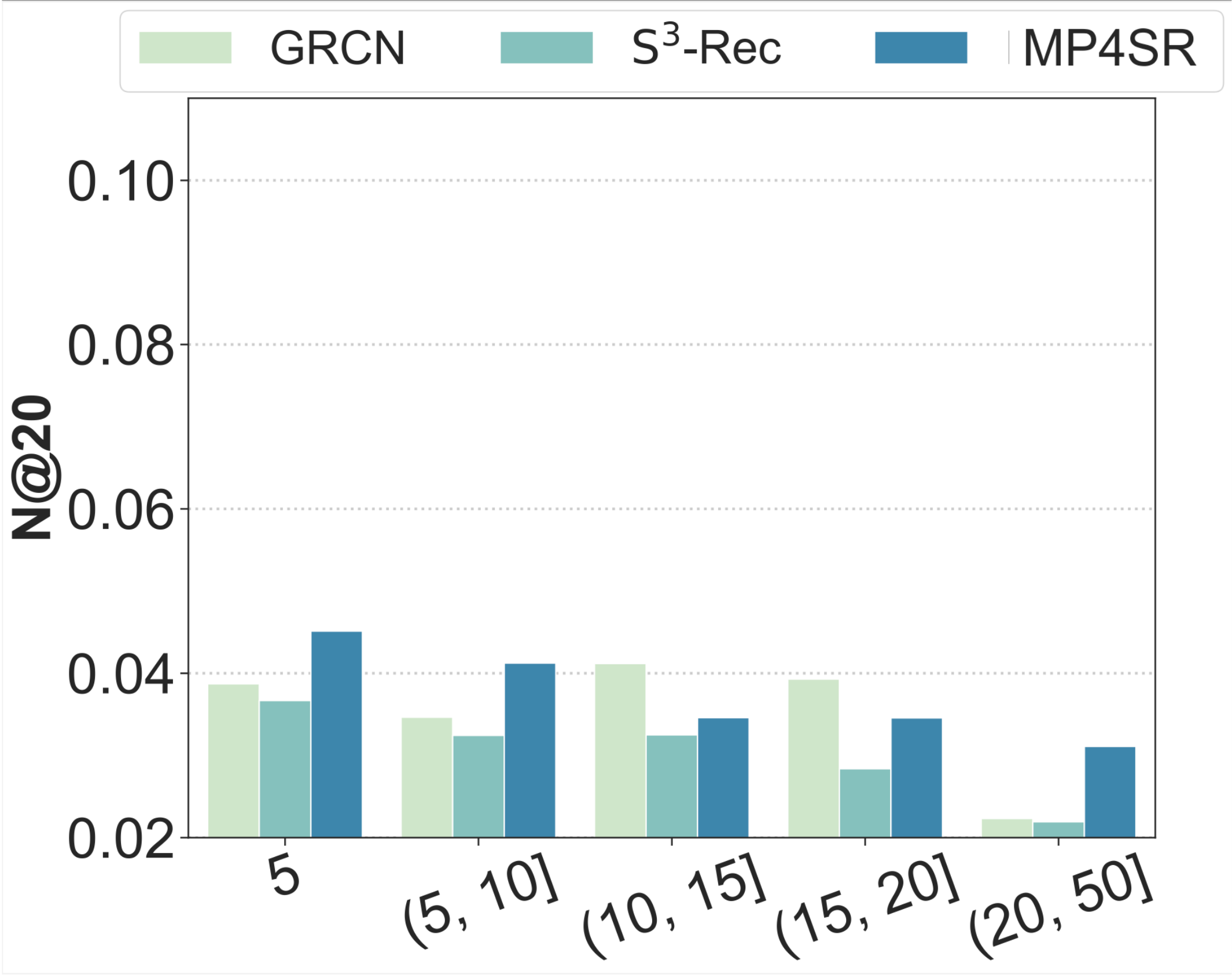}}
\\
\subfloat[Office-Recall]{\includegraphics[width=0.35\textwidth]{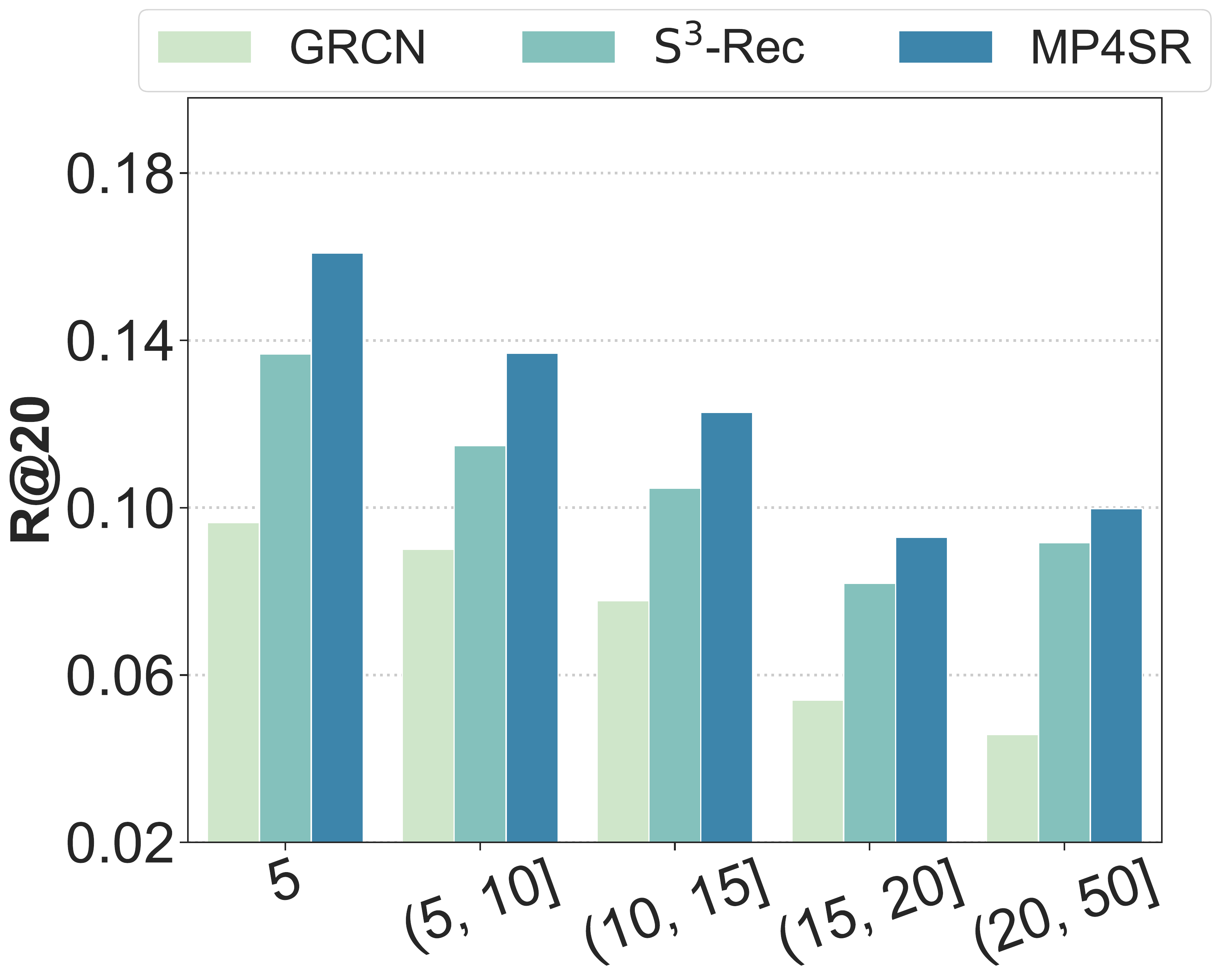}}
\qquad
\subfloat[Office-NDCG]{\includegraphics[width=0.35\textwidth]{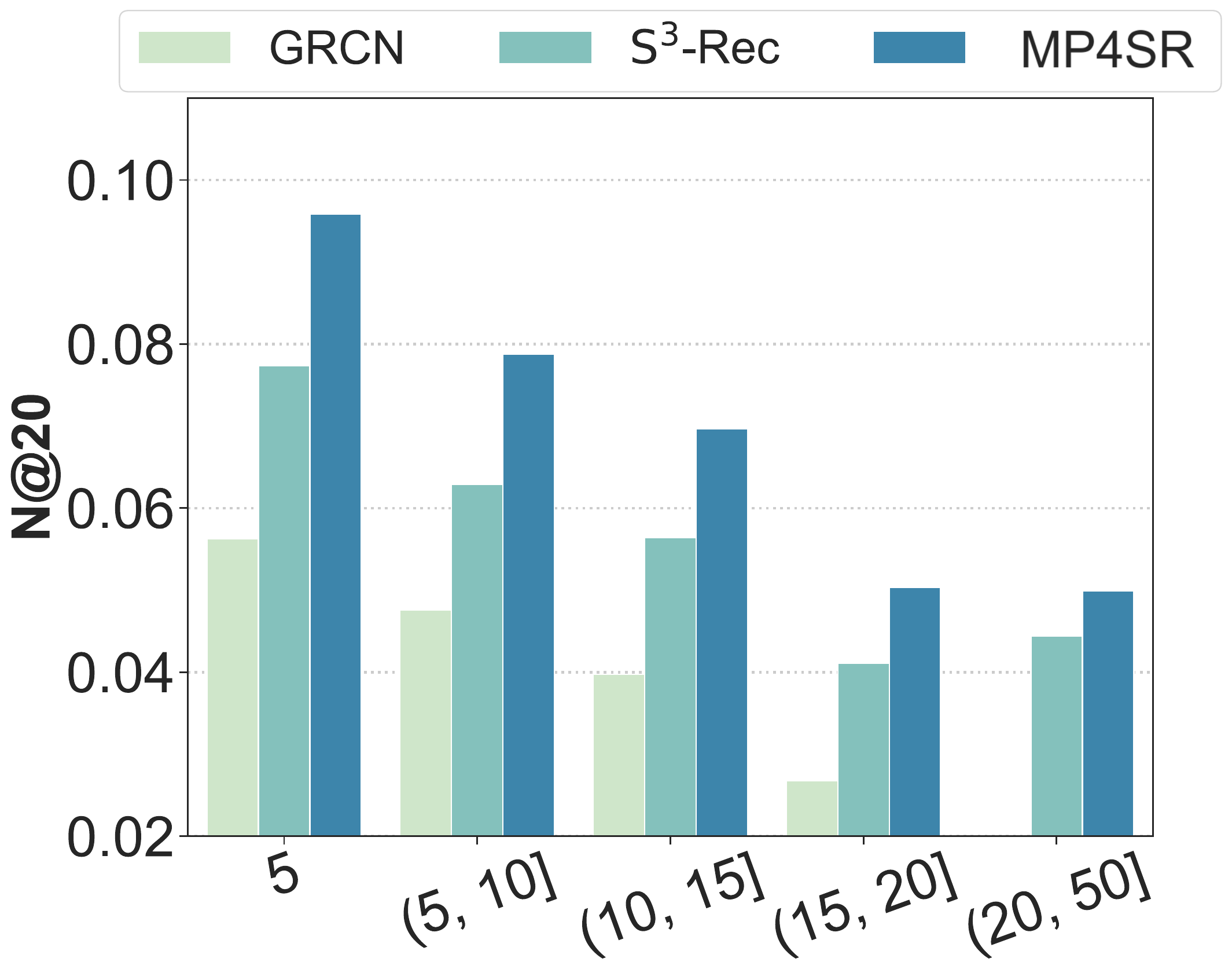}}
\caption{The performance on different user groups achieved by GRCN, S$^3$-Rec, and MP4SR on Pantry and Office datasets.}
\label{fig:perfrom-groups-1}
\end{figure}

The performance comparison results outlined in Table~\ref{tab:overall} allow us to examine the influence of data sparsity on users. Specifically, we split all users into five groups based on the length of their interaction sequences, and assess the performance of the models on each user group. Figure~\ref{fig:perfrom-groups-1} presents a comparison of performance on two datasets, from which we have the following observations.
\textit{Firstly}, for the Office dataset, the proposed MP4SR model performs better than GRCN and $\textrm{S}^3$-Rec across all user groups. When considering the Pantry dataset, both GRCN and MP4SR exceed $\textrm{S}^3$-Rec's performance when user sequences are longer, affirming the importance of multimodal features. \textit{Secondly}, a decrease in the sequence length of user behaviors leads to greater improvements for MP4SR compared to GRCN and $\textrm{S}^3$-Rec. 
This demonstrates the superiority of MP4SR in handling sparse scenarios.

\subsection{\textcolor{black}{Pre-training with Larger Datasets}}
\textcolor{black}{
Works~\cite{radford2021learning,bao2022vlmo, lu2019vilbert,wang2022image} in the field of computer vision have demonstrated the benefits of pre-training on ample data for improving generalization when fine-tuning on downstream datasets for multimodal tasks. Therefore, we aim to investigate whether this conclusion holds true for MP4SR. Specifically, we pre-train the model using all three Amazon datasets (Pantry, Arts, and Office) and then utilize MP4SR for pre-training and fine-tuning the model on each dataset. The results for Pantry and Office are presented in Table~\ref{tab:larger}. 
As anticipated, augmenting the model's pre-training with additional data has indeed improved the sequential recommendation performance for the Pantry dataset. Conversely, for the Office dataset, expanding the pre-training corpus did not yield any enhancement in performance. This discrepancy may be attributed to the Office dataset being the largest among the three datasets employed for pre-training. 
}

\begin{table}
\color{black}
    \caption{\textcolor{black}{The performance of MP4SR which has been trained on a larger dataset.} }
    % \small
    \centering
    \begin{tabular}{l|l|c c| c c}
    \toprule
    Dataset & Model & R@10& R@20 & N@10 & N@20\\
    \midrule
    \multirow{2}{*}{Pantry}
      &MP4SR &0.0673 &0.1040&0.0321&0.0414\\
    &MP4SR\textsubscript{Larger} &\textbf{0.0709} &\textbf{0.1058}&\textbf{0.0334}&\textbf{0.0421}\\
    \midrule
    \multirow{2}{*}{Office}
    &MP4SR &\textbf{0.1206}&\textbf{0.1480}&\textbf{0.0797}&\textbf{0.0866} \\
    &MP4SR\textsubscript{Larger} &0.1195&0.1474&0.0764&0.0834\\
    \bottomrule
    \end{tabular}
    \label{tab:larger}
\end{table}

\begin{table}
    \caption{The performance achieved by MP4SR under unimodal and multimodal-based settings on each dataset. }
    % \small
    \centering
    \begin{tabular}{l|l|c c c c}
    \toprule
    Dataset & Model & R@10 & R@20 & N@10 & N@20\\
    \midrule
    \multirow{3}{*}{Pantry}
    &MP4SR-V&0.0596&0.0928&0.0290&0.0373\\
    &MP4SR-T&0.0649&0.1001&0.0318&0.0407\\
    &MP4SR&\textbf{0.0673}&\textbf{0.1040}&\textbf{0.0321}&\textbf{0.0414}\\
    \midrule
    \multirow{3}{*}{Arts}
    &MP4SR-V&0.1018&0.1345&0.0581&0.0663\\
    &MP4SR-T&0.1144&0.1523&\textbf{0.0652}&\textbf{0.0748}\\
    &MP4SR&\textbf{0.1184}&\textbf{0.1570}&0.0637&0.0735\\
    \midrule
    \multirow{3}{*}{Office}
    &MP4SR-V&0.1153&0.1415&0.0764&0.0830\\
    &MP4SR-T&0.1186&0.1464&0.0772&0.0841\\
    &MP4SR&\textbf{0.1206}&\textbf{0.1480}&\textbf{0.0797}&\textbf{0.0866}\\
    \bottomrule
    \end{tabular}
    \label{tab:compare}
\end{table}

\subsection{Unimodal vs. Multimodal Performance}
\label{sec:unimodal}
MP4SR can be applied when only one modality is available. 
To study the effectiveness of MP4SR in exploiting different modality information, we consider the following two variants of MP4SR for evaluation: 
\begin{itemize}
    \item \textbf{MP4SR-V}: the text modality information is not used for model fine-tuning (\ie removing $\B{h}^{t}(\B{F}^t+\B{E})^{\top}$ from Eq.~(12));
    \item \textbf{MP4SR-T}: the image modality information is not used for model fine-tuning (\ie removing $\B{h}^{v}(\B{F}^v+\B{E})^{\top}$ from Eq.~(12)). 
\end{itemize}
In MP4SR-V, MP4SR-T, and MP4SR, the model is pre-trained with both text and image modalities. 

Table~\ref{tab:compare} presents the recommendation performance of MP4SR-V, MP4SR-T, and MP4SR on each dataset. 
We can note that MP4SR-T outperforms MP4SR-V on all datasets, indicating that text information of items contributes more to performance gain than item images. Leveraging both text and image modality information leads to the best recommendation performance for most datasets. This illustrates the effectiveness of exploiting items' multimodal information for sequential recommendation.

%%%%%%%%%%%%%%%%%%%%%%%%%%%%%%%%%%%%%%%%%%%%%%%%%%%%%%%%%%%%%%%%%%%%%%%%
\section{Conclusion}
\textcolor{black}{This paper presents a novel pre-training framework, called MP4SR (\ie Multimodal Pre-training for Sequential Recommendation), designed with the objective of enhancing sequential recommendation performance through effective integration of multimodal information. The MP4SR framework first transforms item images into textual tokens, a step that serves to reconcile the incongruity between textual and visual modalities. Subsequently, MP4SR employs a backbone network, M$^2$SE (\ie Multimodal Mixup Sequence Encoder), to seamlessly fuse items' multimodal content with the user behavior sequence.}
Two contrastive learning losses are designed to help M$^2$SE learn generalized multimodal sequence representations. 

The experiments on real-world datasets show that the proposed pre-training framework can improve sequential recommendation performance in different settings by effectively regularizing the parameter space for the sequential recommendation. \textcolor{black}{The results demonstrate the successful integration of multimodal pre-training within sequential recommendation representation learning, suggesting that MP4SR is not only an innovative framework but also a valuable contribution to the field of sequential recommendation. As an early attempt in this domain, this paper provides a foundation for the further exploration of multimodal pre-training in sequential recommendation.}

\begin{acks}
This research is supported by Alibaba Group through Alibaba Innovative Research (AIR) Program and Alibaba-NTU Singapore Joint Research Institute (JRI), Nanyang Technological University, Singapore.
\end{acks}

%%
%% The next two lines define the bibliography style to be used, and
%% the bibliography file.
\bibliographystyle{ACM-Reference-Format}
\bibliography{main}

%%% -*-BibTeX-*-
%%% Do NOT edit. File created by BibTeX with style
%%% ACM-Reference-Format-Journals [18-Jan-2012].

\begin{thebibliography}{72}

%%% ====================================================================
%%% NOTE TO THE USER: you can override these defaults by providing
%%% customized versions of any of these macros before the \bibliography
%%% command.  Each of them MUST provide its own final punctuation,
%%% except for \shownote{}, \showDOI{}, and \showURL{}.  The latter two
%%% do not use final punctuation, in order to avoid confusing it with
%%% the Web address.
%%%
%%% To suppress output of a particular field, define its macro to expand
%%% to an empty string, or better, \unskip, like this:
%%%
%%% \newcommand{\showDOI}[1]{\unskip}   % LaTeX syntax
%%%
%%% \def \showDOI #1{\unskip}           % plain TeX syntax
%%%
%%% ====================================================================

\ifx \showCODEN    \undefined \def \showCODEN     #1{\unskip}     \fi
\ifx \showDOI      \undefined \def \showDOI       #1{#1}\fi
\ifx \showISBNx    \undefined \def \showISBNx     #1{\unskip}     \fi
\ifx \showISBNxiii \undefined \def \showISBNxiii  #1{\unskip}     \fi
\ifx \showISSN     \undefined \def \showISSN      #1{\unskip}     \fi
\ifx \showLCCN     \undefined \def \showLCCN      #1{\unskip}     \fi
\ifx \shownote     \undefined \def \shownote      #1{#1}          \fi
\ifx \showarticletitle \undefined \def \showarticletitle #1{#1}   \fi
\ifx \showURL      \undefined \def \showURL       {\relax}        \fi
% The following commands are used for tagged output and should be
% invisible to TeX
\providecommand\bibfield[2]{#2}
\providecommand\bibinfo[2]{#2}
\providecommand\natexlab[1]{#1}
\providecommand\showeprint[2][]{arXiv:#2}

\bibitem[Bao et~al\mbox{.}(2022)]%
        {bao2022vlmo}
\bibfield{author}{\bibinfo{person}{Hangbo Bao}, \bibinfo{person}{Wenhui Wang}, \bibinfo{person}{Li Dong}, \bibinfo{person}{Qiang Liu}, \bibinfo{person}{Owais~Khan Mohammed}, \bibinfo{person}{Kriti Aggarwal}, \bibinfo{person}{Subhojit Som}, \bibinfo{person}{Songhao Piao}, {and} \bibinfo{person}{Furu Wei}.} \bibinfo{year}{2022}\natexlab{}.
\newblock \showarticletitle{Vlmo: Unified vision-language pre-training with mixture-of-modality-experts}.
\newblock \bibinfo{journal}{\emph{Advances in Neural Information Processing Systems}} (\bibinfo{year}{2022}), \bibinfo{pages}{32897--32912}.
\newblock


\bibitem[Chang et~al\mbox{.}(2021)]%
        {chang2021sequential}
\bibfield{author}{\bibinfo{person}{Jianxin Chang}, \bibinfo{person}{Chen Gao}, \bibinfo{person}{Yu Zheng}, \bibinfo{person}{Yiqun Hui}, \bibinfo{person}{Yanan Niu}, \bibinfo{person}{Yang Song}, \bibinfo{person}{Depeng Jin}, {and} \bibinfo{person}{Yong Li}.} \bibinfo{year}{2021}\natexlab{}.
\newblock \showarticletitle{Sequential Recommendation with Graph Neural Networks}. In \bibinfo{booktitle}{\emph{Proceedings of the 44th International ACM SIGIR Conference on Research and Development in Information Retrieval}}. \bibinfo{pages}{378--387}.
\newblock


\bibitem[Cheng et~al\mbox{.}(2021)]%
        {cheng2021learning}
\bibfield{author}{\bibinfo{person}{Mingyue Cheng}, \bibinfo{person}{Fajie Yuan}, \bibinfo{person}{Qi Liu}, \bibinfo{person}{Xin Xin}, {and} \bibinfo{person}{Enhong Chen}.} \bibinfo{year}{2021}\natexlab{}.
\newblock \showarticletitle{Learning transferable user representations with sequential behaviors via contrastive pre-training}. In \bibinfo{booktitle}{\emph{2021 IEEE International Conference on Data Mining (ICDM)}}. IEEE, \bibinfo{pages}{51--60}.
\newblock


\bibitem[Cui et~al\mbox{.}(2018)]%
        {cui2018mv}
\bibfield{author}{\bibinfo{person}{Qiang Cui}, \bibinfo{person}{Shu Wu}, \bibinfo{person}{Qiang Liu}, \bibinfo{person}{Wen Zhong}, {and} \bibinfo{person}{Liang Wang}.} \bibinfo{year}{2018}\natexlab{}.
\newblock \showarticletitle{MV-RNN: A Multi-View Recurrent Neural Network for Sequential Recommendation}.
\newblock \bibinfo{journal}{\emph{IEEE Transactions on Knowledge and Data Engineering}} \bibinfo{volume}{32}, \bibinfo{number}{2} (\bibinfo{year}{2018}), \bibinfo{pages}{317--331}.
\newblock


\bibitem[Donkers et~al\mbox{.}(2017)]%
        {donkers2017sequential}
\bibfield{author}{\bibinfo{person}{Tim Donkers}, \bibinfo{person}{Benedikt Loepp}, {and} \bibinfo{person}{J{\"u}rgen Ziegler}.} \bibinfo{year}{2017}\natexlab{}.
\newblock \showarticletitle{Sequential User-based Recurrent Neural Network Recommendations}. In \bibinfo{booktitle}{\emph{Proceedings of the 11th ACM Conference on Recommender Systems}}. \bibinfo{pages}{152--160}.
\newblock


\bibitem[Erhan et~al\mbox{.}(2010)]%
        {erhan2010does}
\bibfield{author}{\bibinfo{person}{Dumitru Erhan}, \bibinfo{person}{Aaron Courville}, \bibinfo{person}{Yoshua Bengio}, {and} \bibinfo{person}{Pascal Vincent}.} \bibinfo{year}{2010}\natexlab{}.
\newblock \showarticletitle{Why does unsupervised pre-training help deep learning?}. In \bibinfo{booktitle}{\emph{Proceedings of the 13th International Conference on Artificial Intelligence and Statistics}}. \bibinfo{pages}{201--208}.
\newblock


\bibitem[He and McAuley(2016)]%
        {he2016vbpr}
\bibfield{author}{\bibinfo{person}{Ruining He} {and} \bibinfo{person}{Julian McAuley}.} \bibinfo{year}{2016}\natexlab{}.
\newblock \showarticletitle{VBPR: Visual Bayesian Personalized Ranking from Implicit Feedback}. In \bibinfo{booktitle}{\emph{Proceedings of the AAAI Conference on Artificial Intelligence}}.
\newblock


\bibitem[He et~al\mbox{.}(2020)]%
        {he2020lightgcn}
\bibfield{author}{\bibinfo{person}{Xiangnan He}, \bibinfo{person}{Kuan Deng}, \bibinfo{person}{Xiang Wang}, \bibinfo{person}{Yan Li}, \bibinfo{person}{Yongdong Zhang}, {and} \bibinfo{person}{Meng Wang}.} \bibinfo{year}{2020}\natexlab{}.
\newblock \showarticletitle{LightGCN: Simplifying and Powering Graph Convolution Network for Recommendation}. In \bibinfo{booktitle}{\emph{Proceedings of the 43rd International ACM SIGIR Conference on Research and Development in Information Retrieval}}. \bibinfo{pages}{639--648}.
\newblock


\bibitem[Hou et~al\mbox{.}(2022)]%
        {hou2022towards}
\bibfield{author}{\bibinfo{person}{Yupeng Hou}, \bibinfo{person}{Shanlei Mu}, \bibinfo{person}{Wayne~Xin Zhao}, \bibinfo{person}{Yaliang Li}, \bibinfo{person}{Bolin Ding}, {and} \bibinfo{person}{Ji-Rong Wen}.} \bibinfo{year}{2022}\natexlab{}.
\newblock \showarticletitle{Towards Universal Sequence Representation Learning for Recommender Systems}. In \bibinfo{booktitle}{\emph{Proceedings of the 28th ACM SIGKDD Conference on Knowledge Discovery and Data Mining}}. \bibinfo{pages}{585--593}.
\newblock


\bibitem[Hu et~al\mbox{.}(2023)]%
        {hu2023adaptive}
\bibfield{author}{\bibinfo{person}{Hengchang Hu}, \bibinfo{person}{Wei Guo}, \bibinfo{person}{Yong Liu}, {and} \bibinfo{person}{Min-Yen Kan}.} \bibinfo{year}{2023}\natexlab{}.
\newblock \showarticletitle{Adaptive multi-modalities fusion in sequential recommendation systems}. In \bibinfo{booktitle}{\emph{Proceedings of the 32nd ACM International Conference on Information and Knowledge Management}}. \bibinfo{pages}{843--853}.
\newblock


\bibitem[Hu et~al\mbox{.}(2022)]%
        {hu2022memory}
\bibfield{author}{\bibinfo{person}{Yidan Hu}, \bibinfo{person}{Yong Liu}, \bibinfo{person}{Chunyan Miao}, {and} \bibinfo{person}{Yuan Miao}.} \bibinfo{year}{2022}\natexlab{}.
\newblock \showarticletitle{Memory Bank Augmented Long-tail Sequential Recommendation}. In \bibinfo{booktitle}{\emph{Proceedings of the 31st ACM International Conference on Information \& Knowledge Management}}. \bibinfo{pages}{791--801}.
\newblock


\bibitem[Jia et~al\mbox{.}(2021)]%
        {jia2021scaling}
\bibfield{author}{\bibinfo{person}{Chao Jia}, \bibinfo{person}{Yinfei Yang}, \bibinfo{person}{Ye Xia}, \bibinfo{person}{Yi-Ting Chen}, \bibinfo{person}{Zarana Parekh}, \bibinfo{person}{Hieu Pham}, \bibinfo{person}{Quoc Le}, \bibinfo{person}{Yun-Hsuan Sung}, \bibinfo{person}{Zhen Li}, {and} \bibinfo{person}{Tom Duerig}.} \bibinfo{year}{2021}\natexlab{}.
\newblock \showarticletitle{Scaling up visual and vision-language representation learning with noisy text supervision}. In \bibinfo{booktitle}{\emph{International Conference on Machine Learning}}. PMLR, \bibinfo{pages}{4904--4916}.
\newblock


\bibitem[Kang et~al\mbox{.}(2017)]%
        {kang2017visually}
\bibfield{author}{\bibinfo{person}{Wang-Cheng Kang}, \bibinfo{person}{Chen Fang}, \bibinfo{person}{Zhaowen Wang}, {and} \bibinfo{person}{Julian McAuley}.} \bibinfo{year}{2017}\natexlab{}.
\newblock \showarticletitle{Visually-Aware Fashion Recommendation and Design with Generative Image Models}. In \bibinfo{booktitle}{\emph{2017 IEEE International Conference on Data Mining (ICDM)}}. IEEE, \bibinfo{pages}{207--216}.
\newblock


\bibitem[Kang and McAuley(2018)]%
        {kang2018self}
\bibfield{author}{\bibinfo{person}{Wang-Cheng Kang} {and} \bibinfo{person}{Julian McAuley}.} \bibinfo{year}{2018}\natexlab{}.
\newblock \showarticletitle{Self-Attentive Sequential Recommendation}. In \bibinfo{booktitle}{\emph{2018 IEEE International Conference on Data Mining (ICDM)}}. IEEE, \bibinfo{pages}{197--206}.
\newblock


\bibitem[Kim et~al\mbox{.}(2021)]%
        {kim2021vilt}
\bibfield{author}{\bibinfo{person}{Wonjae Kim}, \bibinfo{person}{Bokyung Son}, {and} \bibinfo{person}{Ildoo Kim}.} \bibinfo{year}{2021}\natexlab{}.
\newblock \showarticletitle{Vilt: Vision-and-language transformer without convolution or region supervision}. In \bibinfo{booktitle}{\emph{International Conference on Machine Learning}}. PMLR, \bibinfo{pages}{5583--5594}.
\newblock


\bibitem[Kingma and Ba(2015)]%
        {kingma2014adam}
\bibfield{author}{\bibinfo{person}{Diederik~P Kingma} {and} \bibinfo{person}{Jimmy Ba}.} \bibinfo{year}{2015}\natexlab{}.
\newblock \showarticletitle{Adam: A Method for Stochastic Optimization}. In \bibinfo{booktitle}{\emph{International Conference on Learning Representations}}.
\newblock


\bibitem[Lei et~al\mbox{.}(2021)]%
        {lei2021semi}
\bibfield{author}{\bibinfo{person}{Chenyi Lei}, \bibinfo{person}{Yong Liu}, \bibinfo{person}{Lingzi Zhang}, \bibinfo{person}{Guoxin Wang}, \bibinfo{person}{Haihong Tang}, \bibinfo{person}{Houqiang Li}, {and} \bibinfo{person}{Chunyan Miao}.} \bibinfo{year}{2021}\natexlab{}.
\newblock \showarticletitle{Semi: A sequential multi-modal information transfer network for e-commerce micro-video recommendations}. In \bibinfo{booktitle}{\emph{Proceedings of the 27th ACM SIGKDD Conference on Knowledge Discovery \& Data Mining}}. \bibinfo{pages}{3161--3171}.
\newblock


\bibitem[Liang et~al\mbox{.}(2023)]%
        {liang2023mmmlp}
\bibfield{author}{\bibinfo{person}{Jiahao Liang}, \bibinfo{person}{Xiangyu Zhao}, \bibinfo{person}{Muyang Li}, \bibinfo{person}{Zijian Zhang}, \bibinfo{person}{Wanyu Wang}, \bibinfo{person}{Haochen Liu}, {and} \bibinfo{person}{Zitao Liu}.} \bibinfo{year}{2023}\natexlab{}.
\newblock \showarticletitle{MMMLP: multi-modal multilayer perceptron for sequential recommendations}. In \bibinfo{booktitle}{\emph{Proceedings of the ACM Web Conference 2023}}. \bibinfo{pages}{1109--1117}.
\newblock


\bibitem[Lin et~al\mbox{.}(2022)]%
        {lin2022towards}
\bibfield{author}{\bibinfo{person}{Xudong Lin}, \bibinfo{person}{Simran Tiwari}, \bibinfo{person}{Shiyuan Huang}, \bibinfo{person}{Manling Li}, \bibinfo{person}{Mike~Zheng Shou}, \bibinfo{person}{Heng Ji}, {and} \bibinfo{person}{Shih-Fu Chang}.} \bibinfo{year}{2022}\natexlab{}.
\newblock \showarticletitle{Towards Fast Adaptation of Pretrained Contrastive Models for Multi-channel Video-Language Retrieval}.
\newblock \bibinfo{journal}{\emph{arXiv preprint arXiv:2206.02082}} (\bibinfo{year}{2022}).
\newblock


\bibitem[Liu et~al\mbox{.}(2019b)]%
        {liu2019userdiverse}
\bibfield{author}{\bibinfo{person}{Fan Liu}, \bibinfo{person}{Zhiyong Cheng}, \bibinfo{person}{Changchang Sun}, \bibinfo{person}{Yinglong Wang}, \bibinfo{person}{Liqiang Nie}, {and} \bibinfo{person}{Mohan Kankanhalli}.} \bibinfo{year}{2019}\natexlab{b}.
\newblock \showarticletitle{User Diverse Preference Modeling by Multimodal Attentive Metric Learning}. In \bibinfo{booktitle}{\emph{Proceedings of the 27th ACM International Conference on Multimedia}}. \bibinfo{pages}{1526--1534}.
\newblock


\bibitem[Liu et~al\mbox{.}(2019a)]%
        {liu2019user}
\bibfield{author}{\bibinfo{person}{Shang Liu}, \bibinfo{person}{Zhenzhong Chen}, \bibinfo{person}{Hongyi Liu}, {and} \bibinfo{person}{Xinghai Hu}.} \bibinfo{year}{2019}\natexlab{a}.
\newblock \showarticletitle{User-Video Co-Attention Network for Personalized Micro-video Recommendation}. In \bibinfo{booktitle}{\emph{Proceedings of the Web Conference 2019}}. \bibinfo{pages}{3020--3026}.
\newblock


\bibitem[Liu et~al\mbox{.}(2021b)]%
        {liu2021pre}
\bibfield{author}{\bibinfo{person}{Yong Liu}, \bibinfo{person}{Susen Yang}, \bibinfo{person}{Chenyi Lei}, \bibinfo{person}{Guoxin Wang}, \bibinfo{person}{Haihong Tang}, \bibinfo{person}{Juyong Zhang}, \bibinfo{person}{Aixin Sun}, {and} \bibinfo{person}{Chunyan Miao}.} \bibinfo{year}{2021}\natexlab{b}.
\newblock \showarticletitle{Pre-training Graph Transformer with Multimodal Side Information for Recommendation}. In \bibinfo{booktitle}{\emph{Proceedings of the 29th ACM International Conference on Multimedia}}. \bibinfo{pages}{2853--2861}.
\newblock


\bibitem[Liu et~al\mbox{.}(2021a)]%
        {liu2021augmenting}
\bibfield{author}{\bibinfo{person}{Zhiwei Liu}, \bibinfo{person}{Ziwei Fan}, \bibinfo{person}{Yu Wang}, {and} \bibinfo{person}{Philip~S Yu}.} \bibinfo{year}{2021}\natexlab{a}.
\newblock \showarticletitle{Augmenting Sequential Recommendation with Pseudo-Prior Items via Reversely Pre-training Transformer}. In \bibinfo{booktitle}{\emph{Proceedings of the 44th International ACM SIGIR Conference on Research and Development in Information Retrieval}}. \bibinfo{pages}{1608--1612}.
\newblock


\bibitem[Lu et~al\mbox{.}(2019)]%
        {lu2019vilbert}
\bibfield{author}{\bibinfo{person}{Jiasen Lu}, \bibinfo{person}{Dhruv Batra}, \bibinfo{person}{Devi Parikh}, {and} \bibinfo{person}{Stefan Lee}.} \bibinfo{year}{2019}\natexlab{}.
\newblock \showarticletitle{Vilbert: Pretraining task-agnostic visiolinguistic representations for vision-and-language tasks}.
\newblock \bibinfo{journal}{\emph{Advances in neural information processing systems}}  \bibinfo{volume}{32} (\bibinfo{year}{2019}).
\newblock


\bibitem[Ni et~al\mbox{.}(2019)]%
        {ni2019justifying}
\bibfield{author}{\bibinfo{person}{Jianmo Ni}, \bibinfo{person}{Jiacheng Li}, {and} \bibinfo{person}{Julian McAuley}.} \bibinfo{year}{2019}\natexlab{}.
\newblock \showarticletitle{Justifying Recommendations using Distantly-Labeled Reviews and Fine-Grained Aspects}. In \bibinfo{booktitle}{\emph{Proceedings of the 2019 Conference on Empirical Methods in Natural Language Processing and the 9th International Joint Conference on Natural Language Processing (EMNLP-IJCNLP)}}. \bibinfo{pages}{188--197}.
\newblock


\bibitem[Pan et~al\mbox{.}(2022)]%
        {pan2022multimodal}
\bibfield{author}{\bibinfo{person}{Xingyu Pan}, \bibinfo{person}{Yushuo Chen}, \bibinfo{person}{Changxin Tian}, \bibinfo{person}{Zihan Lin}, \bibinfo{person}{Jinpeng Wang}, \bibinfo{person}{He Hu}, {and} \bibinfo{person}{Wayne~Xin Zhao}.} \bibinfo{year}{2022}\natexlab{}.
\newblock \showarticletitle{Multimodal Meta-Learning for Cold-Start Sequential Recommendation}. In \bibinfo{booktitle}{\emph{Proceedings of the 31st ACM International Conference on Information \& Knowledge Management}}. \bibinfo{pages}{3421--3430}.
\newblock


\bibitem[Paszke et~al\mbox{.}(2019)]%
        {paszke2019pytorch}
\bibfield{author}{\bibinfo{person}{Adam Paszke}, \bibinfo{person}{Sam Gross}, \bibinfo{person}{Francisco Massa}, \bibinfo{person}{Adam Lerer}, \bibinfo{person}{James Bradbury}, \bibinfo{person}{Gregory Chanan}, \bibinfo{person}{Trevor Killeen}, \bibinfo{person}{Zeming Lin}, \bibinfo{person}{Natalia Gimelshein}, \bibinfo{person}{Luca Antiga}, {et~al\mbox{.}}} \bibinfo{year}{2019}\natexlab{}.
\newblock \showarticletitle{PyTorch: An Imperative Style, High-Performance Deep Learning Library}.
\newblock \bibinfo{journal}{\emph{Advances in Neural Information Processing Systems}} (\bibinfo{year}{2019}), \bibinfo{pages}{8026--8037}.
\newblock


\bibitem[Peng et~al\mbox{.}(2021)]%
        {peng2021ham}
\bibfield{author}{\bibinfo{person}{Bo Peng}, \bibinfo{person}{Zhiyun Ren}, \bibinfo{person}{Srinivasan Parthasarathy}, {and} \bibinfo{person}{Xia Ning}.} \bibinfo{year}{2021}\natexlab{}.
\newblock \showarticletitle{HAM: Hybrid Associations Models for Sequential Recommendation}.
\newblock \bibinfo{journal}{\emph{IEEE Transactions on Knowledge and Data Engineering}} (\bibinfo{year}{2021}).
\newblock


\bibitem[Radford et~al\mbox{.}(2021)]%
        {radford2021learning}
\bibfield{author}{\bibinfo{person}{Alec Radford}, \bibinfo{person}{Jong~Wook Kim}, \bibinfo{person}{Chris Hallacy}, \bibinfo{person}{Aditya Ramesh}, \bibinfo{person}{Gabriel Goh}, \bibinfo{person}{Sandhini Agarwal}, \bibinfo{person}{Girish Sastry}, \bibinfo{person}{Amanda Askell}, \bibinfo{person}{Pamela Mishkin}, \bibinfo{person}{Jack Clark}, {et~al\mbox{.}}} \bibinfo{year}{2021}\natexlab{}.
\newblock \showarticletitle{Learning transferable visual models from natural language supervision}. In \bibinfo{booktitle}{\emph{International Conference on Machine Learning}}. \bibinfo{pages}{8748--8763}.
\newblock


\bibitem[Reimers and Gurevych(2019)]%
        {reimers-2019-sentence-bert}
\bibfield{author}{\bibinfo{person}{Nils Reimers} {and} \bibinfo{person}{Iryna Gurevych}.} \bibinfo{year}{2019}\natexlab{}.
\newblock \showarticletitle{Sentence-BERT: Sentence Embeddings using Siamese BERT-Networks}. In \bibinfo{booktitle}{\emph{Proceedings of the 2019 Conference on Empirical Methods in Natural Language Processing and the 9th International Joint Conference on Natural Language Processing (EMNLP-IJCNLP)}}. \bibinfo{pages}{3982--3992}.
\newblock


\bibitem[Rendle et~al\mbox{.}(2010)]%
        {rendle2010factorizing}
\bibfield{author}{\bibinfo{person}{Steffen Rendle}, \bibinfo{person}{Christoph Freudenthaler}, {and} \bibinfo{person}{Lars Schmidt-Thieme}.} \bibinfo{year}{2010}\natexlab{}.
\newblock \showarticletitle{Factorizing Personalized Markov Chains for Next-Basket Recommendation}. In \bibinfo{booktitle}{\emph{Proceedings of the Web Conference 2010}}. \bibinfo{pages}{811--820}.
\newblock


\bibitem[Shazeer et~al\mbox{.}(2017)]%
        {shazeer2017outrageously}
\bibfield{author}{\bibinfo{person}{Noam Shazeer}, \bibinfo{person}{Azalia Mirhoseini}, \bibinfo{person}{Krzysztof Maziarz}, \bibinfo{person}{Andy Davis}, \bibinfo{person}{Quoc Le}, \bibinfo{person}{Geoffrey Hinton}, {and} \bibinfo{person}{Jeff Dean}.} \bibinfo{year}{2017}\natexlab{}.
\newblock \showarticletitle{Outrageously Large Neural Networks: The Sparsely-Gated Mixture-of-Experts Layer}. In \bibinfo{booktitle}{\emph{International Conference on Learning Representations}}.
\newblock


\bibitem[Shuai et~al\mbox{.}(2022)]%
        {shuai2022review}
\bibfield{author}{\bibinfo{person}{Jie Shuai}, \bibinfo{person}{Kun Zhang}, \bibinfo{person}{Le Wu}, \bibinfo{person}{Peijie Sun}, \bibinfo{person}{Richang Hong}, \bibinfo{person}{Meng Wang}, {and} \bibinfo{person}{Yong Li}.} \bibinfo{year}{2022}\natexlab{}.
\newblock \showarticletitle{A review-aware graph contrastive learning framework for recommendation}. In \bibinfo{booktitle}{\emph{Proceedings of the 45th International ACM SIGIR Conference on Research and Development in Information Retrieval}}. \bibinfo{pages}{1283--1293}.
\newblock


\bibitem[Su et~al\mbox{.}(2020)]%
        {su2019vl}
\bibfield{author}{\bibinfo{person}{Weijie Su}, \bibinfo{person}{Xizhou Zhu}, \bibinfo{person}{Yue Cao}, \bibinfo{person}{Bin Li}, \bibinfo{person}{Lewei Lu}, \bibinfo{person}{Furu Wei}, {and} \bibinfo{person}{Jifeng Dai}.} \bibinfo{year}{2020}\natexlab{}.
\newblock \showarticletitle{Vl-bert: Pre-training of generic visual-linguistic representations}. In \bibinfo{booktitle}{\emph{International Conference on Learning Representations}}.
\newblock


\bibitem[Sun et~al\mbox{.}(2019)]%
        {sun2019bert4rec}
\bibfield{author}{\bibinfo{person}{Fei Sun}, \bibinfo{person}{Jun Liu}, \bibinfo{person}{Jian Wu}, \bibinfo{person}{Changhua Pei}, \bibinfo{person}{Xiao Lin}, \bibinfo{person}{Wenwu Ou}, {and} \bibinfo{person}{Peng Jiang}.} \bibinfo{year}{2019}\natexlab{}.
\newblock \showarticletitle{BERT4Rec: Sequential Recommendation with Bidirectional Encoder Representations from Transformer}. In \bibinfo{booktitle}{\emph{Proceedings of the 28th ACM International Conference on Information \& Knowledge Management}}. \bibinfo{pages}{1441--1450}.
\newblock


\bibitem[Tan et~al\mbox{.}(2021)]%
        {tan2021sparse}
\bibfield{author}{\bibinfo{person}{Qiaoyu Tan}, \bibinfo{person}{Jianwei Zhang}, \bibinfo{person}{Jiangchao Yao}, \bibinfo{person}{Ninghao Liu}, \bibinfo{person}{Jingren Zhou}, \bibinfo{person}{Hongxia Yang}, {and} \bibinfo{person}{Xia Hu}.} \bibinfo{year}{2021}\natexlab{}.
\newblock \showarticletitle{Sparse-Interest Network for Sequential Recommendation}. In \bibinfo{booktitle}{\emph{Proceedings of the 14th ACM International Conference on Web Search and Data Mining}}. \bibinfo{pages}{598--606}.
\newblock


\bibitem[Tang and Wang(2018)]%
        {tang2018personalized}
\bibfield{author}{\bibinfo{person}{Jiaxi Tang} {and} \bibinfo{person}{Ke Wang}.} \bibinfo{year}{2018}\natexlab{}.
\newblock \showarticletitle{Personalized Top-N Sequential Recommendation via Convolutional Sequence Embedding}. In \bibinfo{booktitle}{\emph{Proceedings of the 11th ACM International Conference on Web Search and Data Mining}}. \bibinfo{pages}{565--573}.
\newblock


\bibitem[Tolstikhin et~al\mbox{.}(2021)]%
        {tolstikhin2021mlp}
\bibfield{author}{\bibinfo{person}{Ilya~O Tolstikhin}, \bibinfo{person}{Neil Houlsby}, \bibinfo{person}{Alexander Kolesnikov}, \bibinfo{person}{Lucas Beyer}, \bibinfo{person}{Xiaohua Zhai}, \bibinfo{person}{Thomas Unterthiner}, \bibinfo{person}{Jessica Yung}, \bibinfo{person}{Andreas Steiner}, \bibinfo{person}{Daniel Keysers}, \bibinfo{person}{Jakob Uszkoreit}, {et~al\mbox{.}}} \bibinfo{year}{2021}\natexlab{}.
\newblock \showarticletitle{Mlp-mixer: An all-mlp architecture for vision}.
\newblock \bibinfo{journal}{\emph{Advances in neural information processing systems}}  \bibinfo{volume}{34} (\bibinfo{year}{2021}), \bibinfo{pages}{24261--24272}.
\newblock


\bibitem[Vaswani et~al\mbox{.}(2017)]%
        {vaswani2017attention}
\bibfield{author}{\bibinfo{person}{Ashish Vaswani}, \bibinfo{person}{Noam Shazeer}, \bibinfo{person}{Niki Parmar}, \bibinfo{person}{Jakob Uszkoreit}, \bibinfo{person}{Llion Jones}, \bibinfo{person}{Aidan~N Gomez}, \bibinfo{person}{Lukasz Kaiser}, {and} \bibinfo{person}{Illia Polosukhin}.} \bibinfo{year}{2017}\natexlab{}.
\newblock \showarticletitle{Attention is All you Need}.
\newblock \bibinfo{journal}{\emph{Advances in Neural Information Processing Systems}} (\bibinfo{year}{2017}), \bibinfo{pages}{5998--6008}.
\newblock


\bibitem[Wan and McAuley(2018)]%
        {DBLP:conf/recsys/WanM18}
\bibfield{author}{\bibinfo{person}{Mengting Wan} {and} \bibinfo{person}{Julian~J. McAuley}.} \bibinfo{year}{2018}\natexlab{}.
\newblock \showarticletitle{Item recommendation on monotonic behavior chains}. In \bibinfo{booktitle}{\emph{Proceedings of the 12th {ACM} Conference on Recommender Systems, RecSys 2018, Vancouver, BC, Canada, October 2-7, 2018}}, \bibfield{editor}{\bibinfo{person}{Sole Pera}, \bibinfo{person}{Michael~D. Ekstrand}, \bibinfo{person}{Xavier Amatriain}, {and} \bibinfo{person}{John O'Donovan}} (Eds.). \bibinfo{publisher}{{ACM}}, \bibinfo{pages}{86--94}.
\newblock


\bibitem[Wan et~al\mbox{.}(2019)]%
        {DBLP:conf/acl/WanMNM19}
\bibfield{author}{\bibinfo{person}{Mengting Wan}, \bibinfo{person}{Rishabh Misra}, \bibinfo{person}{Ndapa Nakashole}, {and} \bibinfo{person}{Julian~J. McAuley}.} \bibinfo{year}{2019}\natexlab{}.
\newblock \showarticletitle{Fine-Grained Spoiler Detection from Large-Scale Review Corpora}. In \bibinfo{booktitle}{\emph{Proceedings of the 57th Conference of the Association for Computational Linguistics, {ACL} 2019, Florence, Italy, July 28- August 2, 2019, Volume 1: Long Papers}}, \bibfield{editor}{\bibinfo{person}{Anna Korhonen}, \bibinfo{person}{David~R. Traum}, {and} \bibinfo{person}{Llu{\'{\i}}s M{\`{a}}rquez}} (Eds.). \bibinfo{publisher}{Association for Computational Linguistics}, \bibinfo{pages}{2605--2610}.
\newblock


\bibitem[Wang et~al\mbox{.}(2022c)]%
        {wang2022transrec}
\bibfield{author}{\bibinfo{person}{Jie Wang}, \bibinfo{person}{Fajie Yuan}, \bibinfo{person}{Mingyue Cheng}, \bibinfo{person}{Joemon~M Jose}, \bibinfo{person}{Chenyun Yu}, \bibinfo{person}{Beibei Kong}, \bibinfo{person}{Zhijin Wang}, \bibinfo{person}{Bo Hu}, {and} \bibinfo{person}{Zang Li}.} \bibinfo{year}{2022}\natexlab{c}.
\newblock \showarticletitle{TransRec: Learning Transferable Recommendation from Mixture-of-Modality Feedback}.
\newblock \bibinfo{journal}{\emph{arXiv preprint arXiv:2206.06190}} (\bibinfo{year}{2022}).
\newblock


\bibitem[Wang et~al\mbox{.}(2021b)]%
        {wang2021dualgnn}
\bibfield{author}{\bibinfo{person}{Qifan Wang}, \bibinfo{person}{Yinwei Wei}, \bibinfo{person}{Jianhua Yin}, \bibinfo{person}{Jianlong Wu}, \bibinfo{person}{Xuemeng Song}, {and} \bibinfo{person}{Liqiang Nie}.} \bibinfo{year}{2021}\natexlab{b}.
\newblock \showarticletitle{DualGNN: Dual Graph Neural Network for Multimedia Recommendation}.
\newblock \bibinfo{journal}{\emph{IEEE Transactions on Multimedia}} (\bibinfo{year}{2021}).
\newblock


\bibitem[Wang et~al\mbox{.}(2019)]%
        {wang2019sequential}
\bibfield{author}{\bibinfo{person}{S Wang}, \bibinfo{person}{L Hu}, \bibinfo{person}{Y Wang}, \bibinfo{person}{L Cao}, \bibinfo{person}{QZ Sheng}, {and} \bibinfo{person}{M Orgun}.} \bibinfo{year}{2019}\natexlab{}.
\newblock \showarticletitle{Sequential Recommender Systems: Challenges, Progress and Prospects}. In \bibinfo{booktitle}{\emph{Proceedings of the 28th International Joint Conference on Artificial Intelligence}}. \bibinfo{pages}{6332--6338}.
\newblock


\bibitem[Wang et~al\mbox{.}(2022a)]%
        {wang2022image}
\bibfield{author}{\bibinfo{person}{Wenhui Wang}, \bibinfo{person}{Hangbo Bao}, \bibinfo{person}{Li Dong}, \bibinfo{person}{Johan Bjorck}, \bibinfo{person}{Zhiliang Peng}, \bibinfo{person}{Qiang Liu}, \bibinfo{person}{Kriti Aggarwal}, \bibinfo{person}{Owais~Khan Mohammed}, \bibinfo{person}{Saksham Singhal}, \bibinfo{person}{Subhojit Som}, {et~al\mbox{.}}} \bibinfo{year}{2022}\natexlab{a}.
\newblock \showarticletitle{Image as a foreign language: Beit pretraining for all vision and vision-language tasks}.
\newblock \bibinfo{journal}{\emph{arXiv preprint arXiv:2208.10442}} (\bibinfo{year}{2022}).
\newblock


\bibitem[Wang et~al\mbox{.}(2021a)]%
        {wang2021leveraging}
\bibfield{author}{\bibinfo{person}{Xi Wang}, \bibinfo{person}{Iadh Ounis}, {and} \bibinfo{person}{Craig Macdonald}.} \bibinfo{year}{2021}\natexlab{a}.
\newblock \showarticletitle{Leveraging review properties for effective recommendation}. In \bibinfo{booktitle}{\emph{Proceedings of the Web Conference 2021}}. \bibinfo{pages}{2209--2219}.
\newblock


\bibitem[Wang et~al\mbox{.}(2022b)]%
        {wang2021simvlm}
\bibfield{author}{\bibinfo{person}{Zirui Wang}, \bibinfo{person}{Jiahui Yu}, \bibinfo{person}{Adams~Wei Yu}, \bibinfo{person}{Zihang Dai}, \bibinfo{person}{Yulia Tsvetkov}, {and} \bibinfo{person}{Yuan Cao}.} \bibinfo{year}{2022}\natexlab{b}.
\newblock \showarticletitle{Simvlm: Simple visual language model pretraining with weak supervision}. In \bibinfo{booktitle}{\emph{International Conference on Learning Representations}}.
\newblock


\bibitem[Wei et~al\mbox{.}(2021)]%
        {wei2021contrastive}
\bibfield{author}{\bibinfo{person}{Yinwei Wei}, \bibinfo{person}{Xiang Wang}, \bibinfo{person}{Qi Li}, \bibinfo{person}{Liqiang Nie}, \bibinfo{person}{Yan Li}, \bibinfo{person}{Xuanping Li}, {and} \bibinfo{person}{Tat-Seng Chua}.} \bibinfo{year}{2021}\natexlab{}.
\newblock \showarticletitle{Contrastive learning for cold-start recommendation}. In \bibinfo{booktitle}{\emph{Proceedings of the 29th ACM International Conference on Multimedia}}. \bibinfo{pages}{5382--5390}.
\newblock


\bibitem[Wei et~al\mbox{.}(2020)]%
        {wei2020graph}
\bibfield{author}{\bibinfo{person}{Yinwei Wei}, \bibinfo{person}{Xiang Wang}, \bibinfo{person}{Liqiang Nie}, \bibinfo{person}{Xiangnan He}, {and} \bibinfo{person}{Tat-Seng Chua}.} \bibinfo{year}{2020}\natexlab{}.
\newblock \showarticletitle{Graph-Refined Convolutional Network for Multimedia Recommendation with Implicit Feedback}. In \bibinfo{booktitle}{\emph{Proceedings of the 28th ACM International Conference on Multimedia}}. \bibinfo{pages}{3541--3549}.
\newblock


\bibitem[Wei et~al\mbox{.}(2019)]%
        {wei2019mmgcn}
\bibfield{author}{\bibinfo{person}{Yinwei Wei}, \bibinfo{person}{Xiang Wang}, \bibinfo{person}{Liqiang Nie}, \bibinfo{person}{Xiangnan He}, \bibinfo{person}{Richang Hong}, {and} \bibinfo{person}{Tat-Seng Chua}.} \bibinfo{year}{2019}\natexlab{}.
\newblock \showarticletitle{MMGCN: Multi-modal Graph Convolution Network for Personalized Recommendation of Micro-video}. In \bibinfo{booktitle}{\emph{Proceedings of the 27th ACM International Conference on Multimedia}}. \bibinfo{pages}{1437--1445}.
\newblock


\bibitem[Wu et~al\mbox{.}(2022)]%
        {wu2022personalized}
\bibfield{author}{\bibinfo{person}{Yiqing Wu}, \bibinfo{person}{Ruobing Xie}, \bibinfo{person}{Yongchun Zhu}, \bibinfo{person}{Fuzhen Zhuang}, \bibinfo{person}{Xu Zhang}, \bibinfo{person}{Leyu Lin}, {and} \bibinfo{person}{Qing He}.} \bibinfo{year}{2022}\natexlab{}.
\newblock \showarticletitle{Personalized prompts for sequential recommendation}.
\newblock \bibinfo{journal}{\emph{arXiv preprint arXiv:2205.09666}} (\bibinfo{year}{2022}).
\newblock


\bibitem[Xie et~al\mbox{.}(2022a)]%
        {xie2022contrastive}
\bibfield{author}{\bibinfo{person}{Xu Xie}, \bibinfo{person}{Fei Sun}, \bibinfo{person}{Zhaoyang Liu}, \bibinfo{person}{Shiwen Wu}, \bibinfo{person}{Jinyang Gao}, \bibinfo{person}{Jiandong Zhang}, \bibinfo{person}{Bolin Ding}, {and} \bibinfo{person}{Bin Cui}.} \bibinfo{year}{2022}\natexlab{a}.
\newblock \showarticletitle{Contrastive learning for sequential recommendation}. In \bibinfo{booktitle}{\emph{2022 IEEE 38th International Conference on Data Engineering (ICDE)}}. \bibinfo{pages}{1259--1273}.
\newblock


\bibitem[Xie et~al\mbox{.}(2022b)]%
        {xie2022decoupled}
\bibfield{author}{\bibinfo{person}{Yueqi Xie}, \bibinfo{person}{Peilin Zhou}, {and} \bibinfo{person}{Sunghun Kim}.} \bibinfo{year}{2022}\natexlab{b}.
\newblock \showarticletitle{Decoupled Side Information Fusion for Sequential Recommendation}. In \bibinfo{booktitle}{\emph{Proceedings of the 45th International ACM SIGIR Conference on Research and Development in Information Retrieval}}. \bibinfo{pages}{1611–1621}.
\newblock


\bibitem[Xu et~al\mbox{.}(2018)]%
        {xu2018graphcar}
\bibfield{author}{\bibinfo{person}{Qidi Xu}, \bibinfo{person}{Fumin Shen}, \bibinfo{person}{Li Liu}, {and} \bibinfo{person}{Heng~Tao Shen}.} \bibinfo{year}{2018}\natexlab{}.
\newblock \showarticletitle{Graphcar: Content-aware multimedia recommendation with graph autoencoder}. In \bibinfo{booktitle}{\emph{The 41st International ACM SIGIR Conference on Research \& Development in Information Retrieval}}. \bibinfo{pages}{981--984}.
\newblock


\bibitem[Yi and Chen(2021)]%
        {yi2021multi}
\bibfield{author}{\bibinfo{person}{Jing Yi} {and} \bibinfo{person}{Zhenzhong Chen}.} \bibinfo{year}{2021}\natexlab{}.
\newblock \showarticletitle{Multi-Modal Variational Graph Auto-Encoder for Recommendation Systems}.
\newblock \bibinfo{journal}{\emph{IEEE Transactions on Multimedia}}  \bibinfo{volume}{24} (\bibinfo{year}{2021}), \bibinfo{pages}{1067--1079}.
\newblock


\bibitem[Yuan et~al\mbox{.}(2019)]%
        {yuan2019simple}
\bibfield{author}{\bibinfo{person}{Fajie Yuan}, \bibinfo{person}{Alexandros Karatzoglou}, \bibinfo{person}{Ioannis Arapakis}, \bibinfo{person}{Joemon~M Jose}, {and} \bibinfo{person}{Xiangnan He}.} \bibinfo{year}{2019}\natexlab{}.
\newblock \showarticletitle{A Simple Convolutional Generative Network for Next Item Recommendation}. In \bibinfo{booktitle}{\emph{Proceedings of the 12th ACM International Conference on Web Search and Data Mining}}. \bibinfo{pages}{582--590}.
\newblock


\bibitem[Yuan et~al\mbox{.}(2023)]%
        {yuan2023go}
\bibfield{author}{\bibinfo{person}{Zheng Yuan}, \bibinfo{person}{Fajie Yuan}, \bibinfo{person}{Yu Song}, \bibinfo{person}{Youhua Li}, \bibinfo{person}{Junchen Fu}, \bibinfo{person}{Fei Yang}, \bibinfo{person}{Yunzhu Pan}, {and} \bibinfo{person}{Yongxin Ni}.} \bibinfo{year}{2023}\natexlab{}.
\newblock \showarticletitle{Where to go next for recommender systems? id-vs. modality-based recommender models revisited}. In \bibinfo{booktitle}{\emph{Proceedings of the 46th International ACM SIGIR Conference on Research and Development in Information Retrieval}}. \bibinfo{pages}{2639--2649}.
\newblock


\bibitem[Zhang et~al\mbox{.}(2021)]%
        {zhang2021mining}
\bibfield{author}{\bibinfo{person}{Jinghao Zhang}, \bibinfo{person}{Yanqiao Zhu}, \bibinfo{person}{Qiang Liu}, \bibinfo{person}{Shu Wu}, \bibinfo{person}{Shuhui Wang}, {and} \bibinfo{person}{Liang Wang}.} \bibinfo{year}{2021}\natexlab{}.
\newblock \showarticletitle{Mining Latent Structures for Multimedia Recommendation}. In \bibinfo{booktitle}{\emph{Proceedings of the 29th ACM International Conference on Multimedia}}. \bibinfo{pages}{3872--3880}.
\newblock


\bibitem[Zhang et~al\mbox{.}(2022b)]%
        {zhang2022diffusion}
\bibfield{author}{\bibinfo{person}{Lingzi Zhang}, \bibinfo{person}{Yong Liu}, \bibinfo{person}{Xin Zhou}, \bibinfo{person}{Chunyan Miao}, \bibinfo{person}{Guoxin Wang}, {and} \bibinfo{person}{Haihong Tang}.} \bibinfo{year}{2022}\natexlab{b}.
\newblock \showarticletitle{Diffusion-based graph contrastive learning for recommendation with implicit feedback}. In \bibinfo{booktitle}{\emph{International Conference on Database Systems for Advanced Applications}}. Springer, \bibinfo{pages}{232--247}.
\newblock


\bibitem[Zhang et~al\mbox{.}(2024a)]%
        {zhang2024greenrec}
\bibfield{author}{\bibinfo{person}{Lingzi Zhang}, \bibinfo{person}{Yinan Zhang}, \bibinfo{person}{Xin Zhou}, {and} \bibinfo{person}{Zhiqi Shen}.} \bibinfo{year}{2024}\natexlab{a}.
\newblock \showarticletitle{GreenRec: A Large-Scale Dataset for Green Food Recommendation}. In \bibinfo{booktitle}{\emph{Companion Proceedings of the ACM on Web Conference 2024}}. \bibinfo{pages}{625--628}.
\newblock


\bibitem[Zhang et~al\mbox{.}(2024b)]%
        {zhang2024id}
\bibfield{author}{\bibinfo{person}{Lingzi Zhang}, \bibinfo{person}{Xin Zhou}, \bibinfo{person}{Zhiwei Zeng}, {and} \bibinfo{person}{Zhiqi Shen}.} \bibinfo{year}{2024}\natexlab{b}.
\newblock \showarticletitle{Are ID Embeddings Necessary? Whitening Pre-trained Text Embeddings for Effective Sequential Recommendation}. In \bibinfo{booktitle}{\emph{2024 IEEE 40th International Conference on Data Engineering (ICDE)}}.
\newblock


\bibitem[Zhang et~al\mbox{.}(2024c)]%
        {zhang2024dual}
\bibfield{author}{\bibinfo{person}{Lingzi Zhang}, \bibinfo{person}{Xin Zhou}, \bibinfo{person}{Zhiwei Zeng}, {and} \bibinfo{person}{Zhiqi Shen}.} \bibinfo{year}{2024}\natexlab{c}.
\newblock \showarticletitle{Dual-View Whitening on Pre-trained Text Embeddings for Sequential Recommendation}. In \bibinfo{booktitle}{\emph{Proceedings of the AAAI Conference on Artificial Intelligence}}, Vol.~\bibinfo{volume}{38}. \bibinfo{pages}{9332--9340}.
\newblock


\bibitem[Zhang et~al\mbox{.}(2019)]%
        {zhang2019feature}
\bibfield{author}{\bibinfo{person}{Tingting Zhang}, \bibinfo{person}{Pengpeng Zhao}, \bibinfo{person}{Yanchi Liu}, \bibinfo{person}{Victor~S Sheng}, \bibinfo{person}{Jiajie Xu}, \bibinfo{person}{Deqing Wang}, \bibinfo{person}{Guanfeng Liu}, {and} \bibinfo{person}{Xiaofang Zhou}.} \bibinfo{year}{2019}\natexlab{}.
\newblock \showarticletitle{Feature-level Deeper Self-Attention Network for Sequential Recommendation}. In \bibinfo{booktitle}{\emph{Proceedings of the 28th International Joint Conference on Artificial Intelligence}}. \bibinfo{pages}{4320--4326}.
\newblock


\bibitem[Zhang et~al\mbox{.}(2022a)]%
        {IJCAI-GCL4SR}
\bibfield{author}{\bibinfo{person}{Yixin Zhang}, \bibinfo{person}{Yong Liu}, \bibinfo{person}{Yonghui Xu}, \bibinfo{person}{Hao Xiong}, \bibinfo{person}{Chenyi Lei}, \bibinfo{person}{Wei He}, \bibinfo{person}{Lizhen Cui}, {and} \bibinfo{person}{Chunyan Miao}.} \bibinfo{year}{2022}\natexlab{a}.
\newblock \showarticletitle{Enhancing Sequential Recommendation with Graph Contrastive Learning}. In \bibinfo{booktitle}{\emph{Proceedings of the 31st International Joint Conference on Artificial Intelligence}}. \bibinfo{pages}{2398--2405}.
\newblock


\bibitem[Zhao et~al\mbox{.}(2021)]%
        {zhao2021recbole}
\bibfield{author}{\bibinfo{person}{Wayne~Xin Zhao}, \bibinfo{person}{Shanlei Mu}, \bibinfo{person}{Yupeng Hou}, \bibinfo{person}{Zihan Lin}, \bibinfo{person}{Yushuo Chen}, \bibinfo{person}{Xingyu Pan}, \bibinfo{person}{Kaiyuan Li}, \bibinfo{person}{Yujie Lu}, \bibinfo{person}{Hui Wang}, \bibinfo{person}{Changxin Tian}, {et~al\mbox{.}}} \bibinfo{year}{2021}\natexlab{}.
\newblock \showarticletitle{RecBole: Towards a Unified, Comprehensive and Efficient Framework for Recommendation Algorithms}. In \bibinfo{booktitle}{\emph{Proceedings of the 30th ACM International Conference on Information \& Knowledge Management}}. \bibinfo{pages}{4653--4664}.
\newblock


\bibitem[Zhou et~al\mbox{.}(2023b)]%
        {zhou2023comprehensive}
\bibfield{author}{\bibinfo{person}{Hongyu Zhou}, \bibinfo{person}{Xin Zhou}, \bibinfo{person}{Zhiwei Zeng}, \bibinfo{person}{Lingzi Zhang}, {and} \bibinfo{person}{Zhiqi Shen}.} \bibinfo{year}{2023}\natexlab{b}.
\newblock \showarticletitle{A comprehensive survey on multimodal recommender systems: Taxonomy, evaluation, and future directions}.
\newblock \bibinfo{journal}{\emph{arXiv preprint arXiv:2302.04473}} (\bibinfo{year}{2023}).
\newblock


\bibitem[Zhou et~al\mbox{.}(2023c)]%
        {zhou2023enhancing}
\bibfield{author}{\bibinfo{person}{Hongyu Zhou}, \bibinfo{person}{Xin Zhou}, \bibinfo{person}{Lingzi Zhang}, {and} \bibinfo{person}{Zhiqi Shen}.} \bibinfo{year}{2023}\natexlab{c}.
\newblock \showarticletitle{Enhancing dyadic relations with homogeneous graphs for multimodal recommendation}.
\newblock In \bibinfo{booktitle}{\emph{ECAI 2023}}. \bibinfo{publisher}{IOS Press}, \bibinfo{pages}{3123--3130}.
\newblock


\bibitem[Zhou et~al\mbox{.}(2020b)]%
        {zhou2020s3}
\bibfield{author}{\bibinfo{person}{Kun Zhou}, \bibinfo{person}{Hui Wang}, \bibinfo{person}{Wayne~Xin Zhao}, \bibinfo{person}{Yutao Zhu}, \bibinfo{person}{Sirui Wang}, \bibinfo{person}{Fuzheng Zhang}, \bibinfo{person}{Zhongyuan Wang}, {and} \bibinfo{person}{Ji-Rong Wen}.} \bibinfo{year}{2020}\natexlab{b}.
\newblock \showarticletitle{S3-Rec: Self-Supervised Learning for Sequential Recommendation with Mutual Information Maximization}. In \bibinfo{booktitle}{\emph{Proceedings of the 29th ACM International Conference on Information \& Knowledge Management}}. \bibinfo{pages}{1893--1902}.
\newblock


\bibitem[Zhou et~al\mbox{.}(2020a)]%
        {zhou2020unified}
\bibfield{author}{\bibinfo{person}{Luowei Zhou}, \bibinfo{person}{Hamid Palangi}, \bibinfo{person}{Lei Zhang}, \bibinfo{person}{Houdong Hu}, \bibinfo{person}{Jason Corso}, {and} \bibinfo{person}{Jianfeng Gao}.} \bibinfo{year}{2020}\natexlab{a}.
\newblock \showarticletitle{Unified vision-language pre-training for image captioning and vqa}. In \bibinfo{booktitle}{\emph{Proceedings of the AAAI conference on artificial intelligence}}, Vol.~\bibinfo{volume}{34}. \bibinfo{pages}{13041--13049}.
\newblock


\bibitem[Zhou and Shen(2023)]%
        {FREEDOMMM2023}
\bibfield{author}{\bibinfo{person}{Xin Zhou} {and} \bibinfo{person}{Zhiqi Shen}.} \bibinfo{year}{2023}\natexlab{}.
\newblock \showarticletitle{A tale of two graphs: Freezing and denoising graph structures for multimodal recommendation}. In \bibinfo{booktitle}{\emph{Proceedings of the 31st ACM International Conference on Multimedia}}. \bibinfo{pages}{935--943}.
\newblock


\bibitem[Zhou et~al\mbox{.}(2023a)]%
        {BM3WWW2023}
\bibfield{author}{\bibinfo{person}{Xin Zhou}, \bibinfo{person}{Hongyu Zhou}, \bibinfo{person}{Yong Liu}, \bibinfo{person}{Zhiwei Zeng}, \bibinfo{person}{Chunyan Miao}, \bibinfo{person}{Pengwei Wang}, \bibinfo{person}{Yuan You}, {and} \bibinfo{person}{Feijun Jiang}.} \bibinfo{year}{2023}\natexlab{a}.
\newblock \showarticletitle{Bootstrap latent representations for multi-modal recommendation}. In \bibinfo{booktitle}{\emph{Proceedings of the ACM Web Conference 2023}}. \bibinfo{pages}{845--854}.
\newblock


\bibitem[Zhu et~al\mbox{.}(2021)]%
        {zhu2021graph}
\bibfield{author}{\bibinfo{person}{Yanqiao Zhu}, \bibinfo{person}{Yichen Xu}, \bibinfo{person}{Feng Yu}, \bibinfo{person}{Qiang Liu}, \bibinfo{person}{Shu Wu}, {and} \bibinfo{person}{Liang Wang}.} \bibinfo{year}{2021}\natexlab{}.
\newblock \showarticletitle{Graph Contrastive Learning with Adaptive Augmentation}. In \bibinfo{booktitle}{\emph{Proceedings of the Web Conference 2021}}. \bibinfo{pages}{2069--2080}.
\newblock


\end{thebibliography}

\end{document}